\documentclass[11pt]{article}
\usepackage{epsfig}
\usepackage{latexsym} \usepackage{epsf}
\usepackage{a4}
\usepackage{amsfonts}
\usepackage{amsmath}
\usepackage{amsthm}
\usepackage{amssymb}
\renewcommand{\theequation}{\thesection.\arabic{equation}}
\newcounter{subequation}[equation]
\makeatletter

\newcommand{\p}{^{\prime}}

\expandafter\let\expandafter\reset@font\csname reset@font\endcsname

\def\subeqnarray{\arraycolsep1pt
    \def\@eqnnum\stepcounter##1{\stepcounter{subequation}%
        {\reset@font\rm(\theequation\alph{subequation})}}
\jot5mm     \eqnarray}

\makeatother

\def\be{\begin{equation}}
\def\ee{\end{equation}}
\def\bea{\begin{eqnarray}}
\def\eea{\end{eqnarray}}
\def\ba{\begin{array}}
\def\ea{\end{array}}
\def\ep{\varepsilon}
\def\dd{\partial}
\def\o{\omega}
\def\k{\kappa}
\def\half{\frac{1}{2}}

\def\one#1{#1^{\raise5pt\hbox{$\scriptstyle\!\!\!\!1$}}\,{}}
\def\two#1{#1^{\raise5pt\hbox{$\scriptstyle\!\!\!\!2$}}\,{}}

\def\binom#1#2{\left(#1 \atop #2\right)}
\def\tilde{\widetilde}
\def\II{\hbox{{1}\kern-.25em\hbox{l}}}
\def\as{\bar \alpha_S}
\makeatletter
\def\binrel@#1{\begingroup
  \setboxz@h{\thinmuskip0mu
    \medmuskip\m@ne mu\thickmuskip\@ne mu
    \setbox\tw@\hbox{$#1\m@th$}\kern-\wd\tw@
    ${}#1{}\m@th$}%
  \edef\@tempa{\endgroup\let\noexpand\binrel@@
    \ifdim\wdz@<\z@ \mathbin
    \else\ifdim\wdz@>\z@ \mathrel
    \else \relax\fi\fi}%
  \@tempa
}
\let\binrel@@\relax
\def\overset#1#2{\binrel@{#2}%
  \binrel@@{\mathop{\kern\z@#2}\limits^{#1}}}
\def\underset#1#2{\binrel@{#2}%
  \binrel@@{\mathop{\kern\z@#2}\limits_{#1}}}
\makeatother

\begin{document}

\begin{center}
{\LARGE {Small $x$ resummation in 
collinear factorisation  } } \\

\vspace{1cm}

{\large  
R. Kirschner\footnote
{e-mail: Roland.Kirschner@itp.uni-leipzig.de}
  and M. Segond\footnote
{e-mail: Mathieu.Segond@itp.uni-leipzig.de} }

 Institut f\"ur Theoretische
Physik, Universit\"at Leipzig, \\
PF 100 920, D-04009 Leipzig, Germany

\vspace{2cm}
\Large{\it Abstract} \\  
\end{center}

\noindent
The summation of the small $x$ corrections to hard 
scattering QCD amplitudes by collinear factorisation 
method is reconsidered and the K-factor is derived in
leading $\ln x$ approximation with a result differing 
from
the corresponding 
expression by Catani and Hautmann (1994).  
The significance of the difference 
is demonstrated in the examples of structure function 
$F_L$ and of exclusive vector meson electroproduction.
The formulation covers the channels of 
non-vanishing conformal spin $n$ paving the way for  new 
applications.


\section{Introduction}
\setcounter{equation}{0}

Semi-hard processes are characterized by two essentially different large
momentum scales, the hard-scattering scale $Q^2$ and the large c.m.s. energy
squared $s$,  $x$ being the small ratio of theses scales.  
The QCD calculation of the hard
processes involving the factorization of collinear singularities has to be
improved by including  the corrections enhanced by the large logarithm of
$x$. The results of the  QCD Regge asymptotics 
\cite{BFKL} provide the basis
for the resummation  of these large corrections. The method of fitting the
BFKL solution consistently into the collinear factorization, called also 
$k_T$ factorization, has been developed by M. Ciafaloni and collaborators
starting in 1990 \cite{CCH,CH,CC}. 
The idea of $k_T$ factorization and of
unintegrated parton distributions appeared also 
elsewhere (e.g.\cite{Askew:1993jk,Kwiecinski:1995sn,
Kimber:2001sc}),
but the question
of factorization scheme dependence was treated in this work. The scheme has
been worked out and presented in detail in \cite{CH} and has been
reanalyzed in \cite{CC}. 
The resummed  small $x$ corrections affect the hard-scale evolution of the
parton distributions in terms of the anomalous dimension of two-gluon
composite operators and generate a K-factor that can be viewed as an
improvement of the coefficient function.  
Quite a number of papers is relying on this scheme
in general and on the results given in \cite{CH} in particular.  

In the present paper we reconsider the small $x$ resummation.
We follow the known factorization scheme. A peculiar impact 
factor representing the scattering off a parton induces the
collinear singularities. We rely on the factorisation of these singularities in
the small $\ep$ asymptotics. In these details we differ from
the procedure of \cite{CH} and this results in a different 
K-factor.  Our expression has the angular momentum 
singularity  of the BFKL solution. In examples of the structure function
$F_L$ and vector meson electroproduction we demonstrate the significance
of the discrepancy. 

Small $x$ resummation was of great importance for analysing physics at 
HERA and it will be even more important for LHC physics. 
The resummation
has been applied first of all to structure function 
evolution \cite{ASt,ABF,Blumlein:1997em,WT} and to hard inclusive processes like 
heavy quark production \cite{CH,Ball:2001pq}, 
central production
of Drell-Yan pairs \cite{Marzani:2008uh} 
or of Higgs \cite{Marzani:2008az}
and the production 
of prompt photons \cite{Diana:2009xv}.
Recently, the relevance for exclusive semi-hard processes like
electroproduction of vector mesons has been pointed 
out \cite{Ivanov:2007je}.
The resummed K-factor is of particular importance here, because it
cures the prediction instability appearing when going from LO to NLO.
There is no change in the basic scheme when going from the
inclusive case, where forward kinematics applies, to the exclusive
case as long as the momentum transfer remains much smaller than the hard
scale.  Parton distributions have to be substituted by generalized parton distributions (GPD), but still the
BFKL solution in the forward limit applies.

The resummed gluonic anomalous dimension and the K-factor are universal
in the sense, that they do not depend on the details of the process and
also do not necessarily change when going from inclusive to exclusive 
cases. They depend on the exchange channel, merely on the 
quantum number $n$ appearing in the BFKL solution as conformal spin.

The application considered so far concern the channel $n=0$ only. In our
formulation the extension to other values of $n$ is straightforward and
in the following the main steps are done for the general case. In this way
we prepare new applications of the small $x$ resummation, which may concern
both inclusive (e.g. polarized structure functions) and exclusive (e.g.
vector meson electroproduction) cases.

As a general remark we would like to remind that we are discussing the
approximation to perturbative QCD appropriate in the situation if moving
from the Bjorken asymptotics gradually towards to Regge region. 
The logs of the hard scale $Q^2$ are primary and summed first
(eventually with NLO correction). The  logs of  $x$  are
included in the second step as further corrections. 
The applicability is limited
and will be lost if $\ln x$ becomes much larger than $\ln Q^2$.

\subsection{Resummation scheme}

Consider a hard-scattering amplitude or 
structure function calculated in (leading) collinear
approximation, in particular the contribution of
singlet-exchange (vacuum quantum numbers).
\be
A = C_A^{(0)} \otimes GPD, \ \ \
GPD = GPD_0 + P^{(0)} \otimes GPD
\ee
$GPD$ stands for the generalized   parton 
distribution function. The symbol $\otimes$ can be read as
convolution by integrations over longitudinal ($x$)
and transverse momenta ($\k$) or multiplications,  
if double-Mellin representation  
 ($\omega, \gamma$) is used.
 $P^{(0)}$ stands for the DGLAP/ERBL \cite{DGLAP,ERBL} evolution kernel
(or its forward counterpart) and $C_A^{(0)}$ for the
coefficient function.\\

The resummation of the leading 
$\alpha_S \ln \frac{1}{x} \sim \frac{\alpha_S }{\o}$
contribution can be introduced as corrections to the 
coefficient function and to the kernel:
$P^{(0)} \to P, \  C_A^{(0)} \to C_A $.
However, the dominant small $x$ contribution
corresponds to the configuration in the s-channel
intermediate state  where a single two-particle sub-energy
 squared compares to the full energy squared,
$s_{i,i+1} \sim s$. Thus the corrections arise
from the particular iteration loop $i$ only, 
with a sum over $i$.  \\
In the Mellin representation
the small $x$ corrections amount to a 
universal factor $R_\o$ 
and a correction to the leading-order
anomalous dimension $\gamma_{\o}^{(0)}$,
\be 
R_\o = 1+ \sum_{N=1}^{\infty}
(\frac{\alpha_S}{\o})^N r_N, \ \ \  
\gamma_{\o} = \gamma_{\o}^{(0)} + \sum_{N=1}^{\infty}
(\frac{\alpha_S}{\o})^{N+1} b_N, 
\ee
Only the latter is a modification of the gluon
DGLAP kernel.
The first can be regarded as the improvement of
the coefficient function,
\be C_A^{(0)} \to C_A = C_A^{(0)}\, R_\o
\ee
The K-factor $R_\o$ does not depend on the 
kind of scattering particles but merely on the
exchange channel in terms of the BFKL quantum 
number $n$. \\
For calculating $R_\o$ we consider the 
hard-scattering amplitude 
\be A^{(0)} = C_A^{(0)} \otimes GPD^{(0)} \ee
disregarding the DGLAP evolution. Along with
this amplitude we consider the BFKL amplitude
with the same particles as coupling by $C_A^{(0)}$
in high-energy scattering off a parton,
\be A^{BFKL} = \Phi_A \otimes g \otimes \Phi^{part}\;.
\ee
By convolution with a distribution of partons
we obtain an amplitude describing the same scattering
in the small $x$ asymptotics,
\be A^{(x)} = A^{BFKL} \otimes GPD^{(0)} \;.
\ee 
Here $g$ stands for the Green function of BFKL
two-gluon exchange and $\Phi_A$ is the impact factor
coupling the same particles as in the hard 
scattering. $\Phi^{part}$ is the unusual 
partonic impact factor: since its projection onto the channel isotropic in the azimuthal angle is constant in the transverse momentum, it does not obey the 
condition of vanishing with the transverse momenta
which for colourless hadronic impact factors
follows from gauge invariance. 
The convolution of the partonic
impact factor with the BFKL Green function $g$ results in the collinear singularities which are factorized to all order of the coupling constant into an universal transition function $\Gamma (\o,\ep)$,
\bea F^{(0)}&=& g \otimes \Phi^{part} \\ 
&=& F \cdot \Gamma (\o,\ep). 
\eea 
The factor $\Gamma(\o,\ep)$ carrying the collinear
divergencies  is absorbed by redefining the
bare parton distribution $GPD^{(0)}$ 
\be GPD = \Gamma( \o, \ep)\, GPD^{(0)}, \ \ 
 A^{(x)} = \Phi_A \otimes F \otimes GPD \;.\ee
As an essential step for a consistent 
improvement, the factorisation prescription should match the one
adopted in the collinear calculation of the 
hard amplitude, e.g. $\overline {MS}$ scheme. The resulting convolution of $F$ with the parton
distribution, sometimes called unintegrated parton 
gluon distribution, is simply related to the
gluon distribution at small $x$,
\be F\otimes GPD = \gamma_{\o} GPD^{(x)} \;.\ee
In  transverse momenta we expect the form
(no running coupling)
\be F = \gamma_{\o}\, R_\o \,
(\frac{\k^2}{\mu_F^2})^{\gamma_{\o}}.
\ee
We compare now the structure of the original 
hard-scattering amplitude $A$ with the small $x$
amplitude
\be A^{(x)} = \Phi_A \otimes F \otimes GPD^{(x)}. \ee
Identifying the factorisation scale 
with the hard scale 
this results in the wanted improvement of the 
coefficient function,
\be C_A = C_A^{(0)} \, R(\o). \ee 
Simultaneously we learn that in the same 
approximation the impact factor and the 
bare coefficient function are related by
\be C_A^{(0)} = \Phi_A \, \gamma_{\o}. \ee

\section{BFKL in $2+2\ep$ transverse dimensions}
\setcounter{equation}{0}

Consider the BFKL equation in the forward limit in $d=2+2\ep$,
\be
\label{BFKL}
\o \; g(\o,\vec \k, \vec \k_0) =  
\delta^{(2+2\ep)} (\vec \k - \vec \k_0) + 
\frac{\bar \alpha_S}{\mu^{2\ep }} \; \hat K \cdot 
g (\o, \vec \k, \vec \k_0 ) \,.
\ee
 Where we have defined the (bare) dimensionless coupling constant $ \bar \alpha_S = 
\frac{ g^2 N_C}{ 4 \pi^2} $ and    $\mu$ is the fixed scale introduced by dimensional regularisation.
The inhomogeneous term is 
specified in such a way that the solution 
$g (\o, \vec \k, \vec \k_0 )$  is the Green function
of the reggeized two-gluon exchange and the
BFKL amplitude is composed with impact factors
$\Phi_{A/B} $ as
\be
 A^{BFKL} = \int \frac{d^{2+2\ep} \k}{\vec \k^2}
\int \frac{d^{2+2\ep} \k_0}{\vec \k_0^2}
\Phi_A(\vec \k) g (\o, \vec \k, \vec \k_0 )
\Phi_B(\vec \k_0) \;.
\ee
In one-loop approximation the operator $\hat K$ 
acts as 
\be
 \hat K \cdot g (\vec \k, \vec \k_0 ) = 
\frac{1}{\pi} \int
\frac { d^{2+2\ep} \k\p }{(2\pi)^{2 \ep} }
<\vec \k | \hat K |\vec \k\p > g (\vec \k\p, \vec \k_0 )
\ee
with
\be
<\vec \k | \hat K |\vec \k\p > = 
\frac{1}{(\vec \k - \vec \k\p)^2}  -
\half \delta^{(2+2\ep)} (\vec \k - \vec \k\p ) 
 \int \frac{d^{2+2\ep} \k^{\prime \prime} \vec \k^2}
{\vec \k^{\prime \prime \ 2} 
(\vec \k - \vec \k^{\prime \prime} )^2 } \;.
\ee
For solving this equation we shall rely on rotation symmetry 
in $d= 2+2\ep$ dimensions. For $d>2$ representations besides of the 
trivial one have dimension larger that 1 and for
$d>3$ more than one quantum number is needed for
specifying a generic representation. 
We can avoid complications by restricting to the 
class of representations that would 
appear as symmetric traceless tensors of rank $n$.
Instead of working with all spherical harmonics we
can restrict to the ones representing the highest
or lowest weight states in these representations.
This means we consider functions
\be
\label{psi}
\psi_{\gamma , n} (\vec \k) = 
(\vec \k^2 )^{\gamma  -\frac{n}{2}} 
(\vec \epsilon \vec \k)^n, \ \ \  
\psi_{\gamma , n}^{\dagger} (\vec \k) = 
(\vec \k^2 )^{-\gamma  -\frac{n}{2}} 
(\vec \epsilon^* \vec \k)^n
\ee  
where $\vec \epsilon, \vec \epsilon^* $
are two null vectors (with complex-valued components)
such that $ \vec \epsilon \cdot \vec \epsilon^* = 1$ and  $ \vec \epsilon^2 = \vec \epsilon^{* 2}=0$.
They are the $2+2\ep$ dimensional extensions of the 
two-dimensional
vectors $\frac{1}{\sqrt 2}(\vec e_1 \pm i  
\vec e_2 ) $. 
Note that in this paper we only consider positive values of the conformal spin $n$, thus identifying $n$ with $|n|$. \\
In a partial channel of a given $n$ we substitute 
the inhomogeneous term by the projector $\hat \Pi_n$ onto
the highest weight states of representation $n$
and consider the equation which now deals only with dimensionless quantities
\be
\label{BFKLn}
\o  g_n (\o, \vec\k, \vec \k_0) = 
<\vec \k|\hat \Pi_n |\vec \k_0>  + 
\bar \alpha_S^{  r           }                   \;\hat K \cdot  
g_n (\o, \vec\k, \vec \k_0) \;.
\ee
The dimensionless strong coupling which was frozen at $d=2$ dimensions is now running as
\be
\label{alphar}
\bar\alpha_S^{   r  }(\mu_R^2) =\bar \alpha_S \; (\frac{\mu_R^2}{\mu^2})^{\ep}
\ee 
with  the renormalization scale $\mu_R$ and $\bar \alpha_S=\bar\alpha_S^{   r  }(\mu^2)$ is the bare dimensionless strong coupling defined previously.
 We have to calculate 
the action of the kernel on functions 
(\ref{psi}) (see Appendix A)
\be
\label{kgamman}
\hat K \cdot \psi_{\gamma -1 , n} (\vec \k)
 = \lambda (\gamma,n, \ep) \; \psi_{\gamma -1 + \ep , 
n}  \;,
\ee
\be \lambda (\gamma,n, \ep) =  \frac{1}{(4 \pi)^{ \ep}} 
\left [ b(\gamma, n, \ep) - \half b(0, 0, \ep) \; 
\right ],
\ee
\be
b(\gamma, n, \ep) = \Gamma^{-1 } (\ep)
B(\ep, 1+ \frac{n}{2}-\gamma-\ep) \ 
B(\ep, \frac{n}{2} +\gamma + \ep).
\ee
The BFKL kernel can be viewed  as the matrix 
elements of the operator $\hat K$ which can be 
represented in terms
of the quasi-eigenvalues and a shift operator in
$\gamma$.
\be
\label{Kdgamman}
<\gamma,n | \hat K | \gamma_0,n > = \lambda (\gamma_0,n, \ep) \ 
e^{\ep \dd_{\gamma_0} } \ \delta (\gamma- \gamma_0) \,.
\ee
The $\k $ and $\gamma $ representations are related 
by the transition kernels
\be
\label{trans}
<\vec \k|\gamma, n> = \psi_{\gamma , n} (\vec \k), \ \   
<\gamma, n |\vec \k> = 
\psi_{\gamma , n}^{\dagger} (\vec \k) .
\ee
The completeness relation has to be formulated in such a way
that the inhomogeneous term, i.e. $ <\k|\hat \Pi_n |\k\p>$ appears.
\be
 \frac{2}{|S^{(1+2\ep)}|} 
\int \frac{d^{2+2\ep} \k }{(\vec \k^2)^{1+\ep}} 
<\gamma\p, n\p | \vec \k> <\vec \k|\gamma, n > 
= \delta (\gamma\p - \gamma) \delta_{n\p,n} \;, 
\ee
\bea
<\vec \k | \hat \Pi_n |\vec \k\p > &=&\frac{1}{2\pi i} \int_{\half-i\infty}^{\half+i \infty} d\gamma 
<\vec \k| \gamma,n > <\gamma,n | \vec \k\p > \nonumber  \\
&=& 
\delta ( \ln (\frac {{\k\p}^2}{\k^2}) ) (\vec \k^2)^{-n}
(\vec \epsilon \vec \k)^n (\vec \epsilon^* \vec \k\p)^n.
\eea
Here $|S^{(1+2\ep)}| = 2 \pi^{1+\ep} 
\Gamma^{-1}(1+\ep)$ is the area of the unit hypersphere in $d=2+2\ep$ dimensions.\\
We look now on the Green function as 
on matrix elements of the operator
$\hat g_n$,  where $g_n (\vec \k, \vec \k\p) 
=   <\vec \k | \hat g_n | \vec \k\p> $.
The operator equation  reads
\be 
\o \hat g_n = \hat \Pi_n + 
\bar \alpha_S^{r } \; \hat K \cdot \hat g_n 
\ee
and has a simple formal solution. 
In the $\gamma$ representation
where the operator $\hat K $ has the simple form 
(\ref{Kdgamman}) we obtain
\be
<\gamma, n |\hat g_n |\gamma_0, n > = 
\frac{1}{ \o - \bar \alpha_S^{r }\; \lambda (\gamma_0, n, \ep) 
e^{\ep \dd_{\gamma_0}} }
\delta (\gamma - \gamma_0 ). 
\ee
Changing to the original transverse momentum 
representation we obtain
\be
\label{gkkn}
g_n (\k,\k_0) = \frac{1}{2\pi i} \int_{\half-i\infty}^{\half+i\infty} d \gamma
(\k^2 )^{\gamma -\frac{n}{2}} (\vec \epsilon \vec \k )^n       
\frac{1}{ \o - \bar \alpha_S^{r } \  e^{-\ep \dd_{\gamma} } 
\lambda (\gamma, n, \ep) }
(\k_0^2)^{-\gamma - \frac{n}{2}} 
(\vec \epsilon^* \vec \k_0 )^n.
\ee

\section{Factorisation of collinear singularities}
\setcounter{equation}{0}

\subsection{Unintegrated GPD}

The reggeized gluon Green function is not singular in $\ep$,
the singularities appearing in the action 
of the bare kernel and in
the gluon trajectory cancel. 
The collinear singularities appear from the convolution with a parton impact
factor.  No singularity would appear with a hadronic impact factor. 
Therefore, we study the 
convolution with dimensional regularisation 
\bea
\label{F}
F(\o,n, \k,\mu_F,\ep)&=&\int \frac{d^{2 +2\ep}
\k_0 }{\k_0^2 } 
 g_0 (\k, \k_0, \ep ) \theta(\mu^2_F - \k_0^2)
\  \Phi_B (\k_0,\mu,\ep,n)  \nonumber \\
&+&   
\int \frac{d^{2 } \k_0 }{\k_0^2 }
g_0 (\k, \k_0, \ep = 0 )  \theta (\k_0^2 - \mu_F^2) 
 \ 
\Phi_B (\k_0,\mu,0,n) 
\eea 
with the  partonic impact factor 
$\Phi_B (\k_0, \mu,\ep, n) $. In the partial 
channel $n=0$ it is just constant
\be
\label{partonicIF}
\Phi_B (\k_0, \mu,\ep, 0) = 
\frac{\bar\alpha_S }{ \mu^{2\ep}}
\frac{2}{|S^{(1+2\ep)}|} \;    . 
\ee 
This convolution defines the unintegrated gluon 
density or GPD of the parton B  
denoted by $F(\o, \k, n)$ in the following. We shall 
include the case of general $n$ below.
Its convolution with some input hadronic GPD results in 
the unintegrated small-x improved GPD.

For ${\cal R}e \gamma > 0$ divergencies appear in the
first term and they are regularised  by $\ep$ having a 
real part larger than that. The wanted result will be 
obtained by separating  the leading singularity factor.
Less singular contributions, in particular the regular
contribution from the second term (\ref{F}), 
will not affect this result. 
The form of the second term is specified such that
in the double-log approximation
only the term with the leading singular factor 
appears
with no non-leading remainder. This specification is
actually not essential, however the introduction of the
factorisation scale $\mu_F$ in the definition
of a parton impact factor is unavoidable. 

It is known that for hadronic impact factors no
divergencies appear, because they vanish in
the limit $\k_0 \to 0$. In the BFKL Green function the
integration line is ${\cal R}e \gamma = \half$.
The asymptotics in $\k^2$ (leading twist) receives its
contributions from the singularities in the vicinity 
of $\gamma =0$. In the partonic case
the condition of regularisation 
${\cal R}e (\ep - \gamma) > 0$ has to be preserved
and therefore the pole $ (\ep - \gamma)^{-1}$ 
arising fron the integration over $\k_0$
is to be kept to the right of the contour.
We consider the collinear divergent part 
\bea
\label{F<}
F_< (\o, \k) &=&  \frac{\bar\alpha_S }{ \mu^{2\ep}}
\int \frac{d \gamma}{2\pi i}
(\k^2 )^{\gamma} \int_0^{\mu_F^2} d \k_0^2  
\frac{1}{ \o - \bar \alpha_S^r  \, 
e^{-\ep \dd_{\gamma} } \lambda (\gamma, 0, \ep) }
(\k_0^2)^{\ep -\gamma -1} \nonumber  \\  
&=& \frac{\bar \alpha_S}{\o} (\frac{\mu_F^2}{\mu^2})^{\ep}
\int \frac{d \gamma}{2\pi i}
(\frac{\k^2}{\mu_F^2} )^{\gamma}   
\frac{1}{ 1 - \frac{\bar \alpha_S^r}{\o}  
\frac{1}{\gamma} \lambda_1 (\gamma,0,\ep) \,
e^{-\ep \dd_{\gamma} } }
\frac{1}{\ep - \gamma} \;.
\eea 
Owing to the explicit form of 
$\lambda (\gamma, \ep) $ given in Appendix A 
eq.(\ref{lambdagam}) we have substituted it by 
$\frac{1}{\gamma + \ep} \lambda_1 (\gamma + \ep, \ep)$. An important remark has to be done at this point: as we discussed previously, the strong coupling $\bar \alpha_S^r $ in the denominator comes from the BFKL Green function, therefore runs with the yet unknown renormalization scale $\mu_R$. But we explicitly show in the Appendix B how this scale dependance actually disappears, hence we replace from now  $\bar \alpha_S^r$ by $\bar \alpha_S$, and postpone the discussion on fixing $\mu_R$ at the end of the calculation.

We first study the simplified version of $F(\o, \k,0)$ where we put just
$\lambda_1 =1 $ for explaining the
essential steps, denoting the result by $F_{0}(\o, \k)$.
This corresponds to the double-logarithmic approximation.
Indeed, the dependence on $(\gamma,\o) $ enters  essentially 
as $\frac{\bar \alpha_S}{\gamma \,\o }$.
The leading contribution at large $\k^2$ resulting in the large
$Q^2$ asymptotics after convolution with the impact factor involving the
virtual photon, is obtained from the residue in the pole
$ \gamma = \hat \gamma_{\o}^{(0)},  $ with 
\be
\hat \gamma_{\o}^{(0)} = \frac {\bar \alpha_S}{\o} 
e^{-\ep \dd_{\gamma}} \;,
\ee 
as
\be
\label{F0<}
 F_{0 <} (\o, \k, 0) =  
 (\frac{\bar \alpha_S}{\mu^2})^{\ep} \left. 
\int \frac{d \gamma}{2\pi i}
(\k^2 )^{\gamma} 
\int_0^{\mu_F^2} d \k_0^2   
\frac{1}{ \gamma -  \hat \gamma_{\o}^{(0)} } \
\frac{\gamma }{\o}  
(\k_0^2)^{\ep -\gamma   -1} \ 
\right |_{\gamma\p = 0}.
\ee
The factor involving the shift operator can be expanded in 
geometric series and then in each term the shift operator is 
moved to one side.
The calculation is given in  Appendix C, where
it is shown that the result is
\be
 F_{0 <} (\o, \k, 0)
 =  \bar \alpha_S (\frac{\mu_F^2}{\mu^2})^{\ep}
(\frac{\k^2}{\mu_F^2 })^{ \gamma_{\o}^{(0)}}   
\frac{1}{\o} 
\left ( \exp (\frac{1}{\ep} \gamma_{\o}^{(0)} ) - 1  \right ) \;.
\ee
Also an alternative way is described in   Appendix C because it 
is convenient for treating the general case $\lambda_1 \not = 1$. 
The integral over $\gamma$ can be done before operator ordering. 
To avoid interference with the action of the shift operator we replace 
in the integrand $\gamma$  by $\gamma + \gamma\p$,  let the shift operator
act on $\gamma\p$ only and put finally $\gamma\p = 0 $. \\
We add the contribution from $\k^2_0 > \mu_F^2$,
\be
 F_{0 >} (\o, \k, 0) =  \bar \alpha_S
\int \frac{d \gamma}{2\pi i}
(\k^2 )^{\gamma} 
\int_{\mu_F^2}^{\infty} d \k_0^2   
\frac{1}{ \gamma -   \gamma_{\o}^{(0)} } \
\frac{\gamma  }{\o}  
(\k_0^2)^{ -\gamma -1} \ 
= \frac{\bar \alpha_S}{\o} (\frac{\k^2 }{\mu_F^2})^{\gamma_{\o}^{(0)}} 
\ee
and obtain
\be
\label{F0}
F_{0} (\o, \k, 0) =
(\frac{\k^2}{\mu_F^2 })^{ \gamma_{\o}^{(0)}}   
\frac{\bar \alpha_S}{\o} 
 \exp (\frac{1}{\ep} \gamma_{\o}^{(0)} ). 
\ee
Our way of factorizing the collinear singularities
in the convolution of the BFKL Green function with
the partonic impact factor follows the
standard BFKL method and differs technically 
from the
ways followed in \cite{CH,CC}. In order to show
that the reason for the discrepancy in the results
is not in these technical details we treat also the
form  given in \cite{CH} as the first solution. 
This solution of BFKL equation is obtained 
by iteration and 
has been presented by Catani and Hautmann 
in the form
\be
\label{CHdisc}
F
= \sum_{k=0}^{\infty} \left ( 
\frac{\alpha_S}{\omega } C_{\ep} 
( \frac{k^2}{\mu^2} )^{\ep}
\right )^{k+1} C_{k+1}(\ep).
\ee
In \cite{CH}  the iteration starts from a Born 
term representing what we have called parton impact 
factor, so the notation $F$ refers to the same quantity as above.  
$C_{\ep}$ is to specify  the $\overline {MS}$ 
prescription.
We shall use temporarily the abbreviation
$ z_{\ep} = C_{\ep} ( \frac{k^2}{\mu^2} )^{\ep}.
$
The BFKL equation results in the iterative relation for 
the coefficients,
\be
C_{k+1}(\ep) = C_k(\ep) I_k (\ep) \;,
\ee 
with $C_1(\ep) = 1$. The notation $I_k(\ep) $ in \cite{CH} is related to
ours as $ I_k(\ep) = I(k\ep, \ep)$ and 
\be
I(\gamma, \ep) = 
\frac{\lambda_1(\gamma,0, \ep)}{\gamma}
= \lambda(\gamma-\ep, 0, \ep).
\ee 
Note that only the case $n=0$ was  considered 
in \cite{CH} and we restrict ourselves to this case
now. With the help of the shift operator
$T_{\ep} = e^{\ep \dd_{\gamma}\p} $
the coefficients can be written compactly,
\be
 C_{k+1}(\ep) = \prod_{j=1}^k   
\frac{\lambda_1(j \ep, \ep)}{j \ep} =
\left (T_{\ep} \left.
\frac{\lambda_1(\gamma\p , \ep)}{\gamma\p}
\right )^k \right |_{\gamma\p = 0}.
\ee
Defining $\hat \gamma^{(0)}_{\omega} = 
\frac{\alpha_S}{\omega} e^{\ep \dd_{\gamma\p}}$ this allows to do the sum
\be
\label{CHprime}
F = \left. \left ( 
\frac{\alpha_S}{\omega} z_{\ep} 
\right )
\frac{1}{ 1- \hat \gamma^{(0)}_{\omega} z_{\ep}
\frac{\lambda_1(\gamma\p, \ep)}{\gamma\p} }
\right |_{\gamma\p =0} \;.
\ee 
The  double log case where 
$\lambda_1 (\gamma,\ep) = 1$ is easily done
\bea
\label{CHeps}
\left. \frac{1}{ 1- \hat \gamma^{(0)}_{\omega} z_{\ep}
\frac{1}{\gamma} }\right |_{\gamma =0} 
&=& \sum_{N=0}^{\infty} z_{\ep}^N 
 \hat \gamma^{(0)}_{\omega} \frac{1}{\gamma}
 \hat \gamma^{(0)}_{\omega} \frac{1}{\gamma}
...
 \hat \gamma^{(0)}_{\omega} \frac{1}{\gamma}
\left. \right |_{\gamma=0} = \exp ( \frac{1}{\ep} \gamma_{\omega}^{(0)} 
z_{\ep} ) \nonumber \\
&=& \left (\frac{k^2}{\mu^2} \right )^
{\gamma_{\omega}^{(0)} }
e^{\frac{1}{\ep} \gamma_{\omega}^{(0)} }.
\eea 
The result for this simplified case is equivalent to (\ref{F0}) as expected, beside on the $\mu_F$ dependance which will be further discussed in details  for the general case. The details of an alternative method, which starts 
from implementing the condition $\gamma = 0$ by a contour integral are also given in Appendix C.

\subsection{Asymptotics in $\ep$}

We consider now $F_<(\o, \k) $ (\ref{F<}) without omitting 
$\lambda_1 (\gamma, \ep )$. 
We substitute the integration variable $\gamma$ by
\be
\label{tilde}
 \tilde \gamma = \frac{\gamma}
{\lambda_1 (\gamma, \ep) } 
\ee
and obtain
\be
\label{F<tilde}
F_<(\o, \k) = 
\frac{\bar\alpha_S }{\o}
(\frac{ \mu_F^2}{\mu^2})^{\ep} \int 
\frac {d \tilde \gamma}{2\pi i}
\frac{\lambda_1 (\gamma, \ep)}{1- \tilde \gamma   
\lambda\p_1 (\gamma, \ep)}
(\frac{\k^2}{\mu_F^2} )^{\gamma}   
\frac{1}{ \tilde \gamma -  \hat \gamma_{\o}^{(0)}}
 \tilde \gamma \ \frac{1}{\ep -\gamma }  \,.
\ee
Now $\gamma$ is to be considered as a function of 
$\tilde \gamma$ and
$\lambda\p_1 (\gamma, \ep) = \dd_{\gamma} \lambda_1 (\gamma, \ep) $. 
We proceed in close analogy to the above simplified case
treated in detail in  Appendix C. 
 We  do  first the 
$\gamma $ integral by residue following the second 
calculation in  Appendix C.     
The singularities originate from the last factor 
in (\ref{F<tilde}) and it is
sufficient to account for the shift operator action
on this particular factor. We have to  release the integration
variable from its role as the operator conjugate to
the infinitesimal shift. This we do by 
   substituting $\gamma $ by 
$\gamma + \gamma\p$ in the last   pole factor 
and redefining 
 that now $\hat \gamma_{\o}^{(0)}$ is acting on 
$\gamma \p$. The corresponding substitution in the 
remaining factors is unimportant, i.e. amounts to
corrections non-leading in $\ep \to 0$.\\
 We pick up the residue located at the pole 
$
 {\tilde \gamma} = \hat \gamma_{\o}^{(0)}
$
 equivalent to the operatorial equation
\be
\label{gammao}
 \hat \gamma_{\o}(\ep) = \lambda_1 \left(\hat \gamma_{\o}(\ep), \ep \right) \hat \gamma_{\o}^{(0)} \; \; ,
\ee
whose solution  $\hat \gamma_{\o}(\ep)$ can be 
expanded in powers of $\hat \gamma_{\o}^{(0)}$ as shown in  Appendix B  
eq.(\ref{gammasol}). A crucial feature of our approach is that the location of this pole, or equivalently the explicit expression of $\hat \gamma_{\o}(\ep)$ is not affected by the running of the strong coupling as we show in the Appendix B, therefore justifying the replacement done previously  of $\bar\alpha_S^r$ by $\bar \alpha_S$, also hidden in $\hat \gamma_{\o}^{(0)}$. Using the relation (\ref{gammao}), we get
\bea 
F_< (\o, \k)
\label{defAB}
&=&
\left.
\frac{\bar\alpha_S}{\o} (\frac{ \mu_F^2}{\mu^2})^{\ep}
\frac{1}
{1- \hat \gamma_{\o}^{(0)} 
\lambda\p_1 (\hat \gamma_{\o}, \ep) }   
(\frac{\k^2}{\mu_F^2} )^{\hat \gamma_{\o}}   
\hat \gamma_{\o} \frac{1}{-  \gamma\p }
\frac{1}{1 - \hat \gamma_{\o}\frac{1}{- \gamma\p} }
 \right |_{\gamma\p =0} \\ \nonumber
&=& \frac{\as}{\o} (\frac{ \mu_F^2}{\mu^2})^{\ep} \ A(\hat \gamma_{\o}^{(0)}, \ep) (\frac{\k^2}{\mu_F^2} )^{\hat \gamma_{\o}}
B ( \hat \gamma_{\o}^{(0)}, \ep, \gamma\p) \ |_{\gamma\p =0} \;. 
\eea
Now we  evaluate the product
\be 
\label{exp}
 \left. A(\hat \gamma_{\o}^{(0)}, \ep) B ( \hat \gamma_{\o}^{(0)}, \ep, \gamma\p) \right |_{\gamma\p =0} 
= \left. A(\hat \gamma_{\o}^{(0)}, \ep) \sum_{N=1}^{\infty}
\left( \hat \gamma_{\o} \frac{1}{-  \gamma\p } \right)^N
 \right |_{\gamma\p =0} .
\ee 
We write (for $\gamma\p < 0 $)
$ \frac{1}{-\gamma\p} = \int_0^1 
\frac{d\alpha}{\alpha} \alpha^{-\gamma\p}
$ and notice that
\be
\label{alphagamma}
 f (\hat \gamma_{\o}^{(0)}) \alpha^{-\gamma\p}  =
\alpha^{-\gamma\p} \ \  f ( \hat \gamma_{\o}^{(0)} 
\alpha^{\ep} ).
\ee
Therefore we have 
\be  
( \hat \gamma_{\o} \frac{1}{-  \gamma\p } )^N 
= \prod_{i=1}^N  \int_0^1 \frac{d\alpha_i }{\alpha_i} 
\alpha_i^{-\gamma\p} \ \  \prod_{i=1}^N \hat
\gamma_{\o} (\hat \gamma_{\o}^{(0)}
\alpha_1^{\ep}... \alpha_{i}^{\ep},\ep ). 
\ee 
We substitute $ \beta_{i} =  \alpha_1^{\ep} \alpha_{2}^{\ep} ...
\alpha_i^{\ep} $ and obtain 
\bea
A(\hat \gamma_{\o}^{(0)}, \ep) B ( \hat \gamma_{\o}^{(0)}, \ep, \gamma\p) \
&=& \sum_{N=1}^{\infty} 
\frac{1}{\ep^N} \int_0^1 \frac{d\beta_1 }{\beta_1}  
\int_0^{\beta_1} \frac{d\beta_2 }{\beta_2}  ... 
\int_0^{\beta_{N-1}}  \frac{d\beta_N }{\beta_N} 
\ \ \beta_N^{- \frac{\gamma\p}{ \ep}}  \nonumber
\\  
&\times& \prod_{i=1}^N \hat \gamma_{\o} (\hat \gamma_{\o}^{(0)}\beta_i,\ep ) 
A(\hat \gamma_{\o}^{(0)} \beta_N, \ep).
\eea
After having acted on the whole $\gamma\p$ 
dependent terms, all the involved shift operator 
are  moved to the right and act  on a 
function constant in $\gamma\p$; 
then the shift operators  can be 
substituted by unit operator. We can now do the limit
$ \gamma\p \rightarrow 0^-$.  This means to replace 
\be
 \hat \gamma_{\o}^{(0)} \rightarrow
 \gamma_{\o}^{(0)}=\frac{\bar\alpha_S}{\o}
\ee
and also $\hat \gamma_{\o}(\ep) \rightarrow \gamma_{\o}(\ep)$ 
where $\gamma_{\o}(\ep)$ is the solution of 
\be
  \gamma_{\o}(\ep) = \lambda_1 ( \gamma_{\o}(\ep), \ep)  \gamma_{\o}^{(0)} \,.
\ee
 After this we get
\be
\label{Fresult1}
\left.A(\hat \gamma_{\o}^{(0)}, \ep) B ( \hat \gamma_{\o}^{(0)}, \ep, \gamma\p)\right |_{\gamma\p =0} = \frac{1}{\ep}\int_0^1  \frac{d\beta}{\beta}
\gamma_{\omega} (\gamma_{\omega}^{(0)} \beta, \ep) A( \gamma_{\o}^{(0)} \beta , \ep)
 \exp \left(\frac{1}{\ep} \int_{\beta}^1  
\frac{d\beta_1}{\beta_1} 
\gamma_{\omega} (\gamma_{\omega}^{(0)} \beta_1
,\ep ) \right) \; .
\ee
We compute in the  Appendix D  the asymptotics of this product and we get  
\be
\label{F<result}
F_<(\o,\k) =  
\frac{\bar\alpha_S}{\o} 
\frac{1}
{1-  \gamma_{\o}^{(0)} 
\lambda\p_1 ( \gamma_{\o}, 0) }   
(\frac{\k^2}{\mu_F^2} )^{ \gamma_{\o} }   
\left \{ \exp \left (\frac{1}{\ep} \int_0^1 
\frac{d\alpha}{\alpha} 
\gamma_{\o} (\gamma_{\o}^{(0)} \alpha, \ep ) 
\right ) -1 \right\} \;.
\ee
Outside the exponential $\gamma_{\o}(\ep) $ 
is evaluated at $\ep=0$. The subtraction term in the bracket is actually 
overestimating accuracy. The latter as well as the 
contribution
$F_>$ are non-leading in the asymptotics $\ep \to 0$ and cancel  each other. We rewrite the 
preexponential
factors in terms of the usual  BFKL eigenvalue function (\ref{chi}) in 2 dimensions
$\chi (\gamma)$, using the relation (\ref{derlambda1}),
and  obtain the final result by restoring the explicit 
expression $\bar\alpha_S (\frac{\mu_R^2}{\mu^2})^\ep$ of the dimensionless 
strong coupling  in the argument of the 
anomalous dimension appearing in the exponential term since it leads 
to a factor contributing to the order $\ep^0$, and also with the factor 
$S_{\ep} = \exp \{ -\ep [ \psi (1) + \ln 4 \pi ]\} $ which characterizes 
the $\overline{\mbox {\rm MS}}$-scheme, as\\
\be
\label{Fresult}
 F(\o,\k) =   \gamma_{\o}
\frac{1}
{ -  \gamma_{\o}^2 \chi\p(\gamma_{\o}) }   
(\frac{\k^2}{\mu_F^2} )^{ \gamma_{\o} }   
\exp \left (\frac{1}{\ep} \int_0^{S_{\ep}} \frac{d\alpha}{\alpha} 
\gamma_{\o} (\gamma_{\o}^{(0)}  (\frac{\mu_R^2}{\mu^2})^\ep  \alpha, \ep ) 
\right )
\ee
with 
\be
 \gamma_{\o}(\gamma_{\o}^{(0)},\ep)=\gamma_{\o} (\gamma_{\o}^{(0)})+ \ep \, \gamma_{\o}^{(\ep)}(\gamma_{\o}^{(0)}) + {\cal O} (\ep^2) \;
\ee
where the  explicit expressions of the BFKL gluon anomalous dimension $\gamma_{\o}$ and of its $\ep$-correction $\gamma_{\o}^{(\ep)}$ are given in Appendix B, eq.(\ref{BFKLanomdim}) and (\ref{BFKLanomdimeps}).\\

The analysis of the $\ep$ asymptotics works  with
modification also for the discrete sum solution
(\ref{CHdisc}). We show that this solution 
results in the same asymptotics.
We follow the steps of the double log calculation
as in  Appendix C.
 We include the
 factor $z_{\ep}$ into $\lambda_1 $. 
We use the notation $ \tilde \gamma (\gamma) =
\frac{\gamma}{\lambda_1 (\gamma)} $.
\bea
 \left. \frac{1}{1- \hat \gamma_{\omega}^{(0)} 
\frac{\lambda_1 (\gamma,\ep)}{\gamma} }  
\right|_{\gamma = 0} &=& \left.
\frac{1}{2\pi i} \oint_{C_0} \frac{d\gamma}{\gamma} 
\frac{1}{1- \hat \gamma_{\omega}^{(0)\prime } 
\frac{1}{\tilde \gamma (\gamma + \gamma\p)} }  
\right|_{\gamma\p = 0} \nonumber \\
&=& \left. 
1+ \frac{1}{2\pi i} \oint_{C_0} \frac{d\gamma}{\gamma} 
\hat \gamma_{\omega}^{(0) \prime}
\frac{1}{\tilde \gamma (\gamma + \gamma\p)- 
\hat \gamma_{\omega}^{(0) \prime}  } 
\right|_{\gamma\p = 0} \nonumber \\
&=& \  
\left.
1+ \frac{1}{2\pi i} \oint_{\hat C} 
\frac{d\gamma^{\prime \prime}}
{\gamma^{\prime \prime} -\gamma\p} 
\frac{1}{\tilde \gamma (\gamma^{\prime \prime} )- 
\hat \gamma_{\omega}^{(0) \prime T}  }
\hat \gamma_{\omega}^{(0) \prime T} 
\right|_{\gamma\p = 0} \;.
\eea
We change the integration variable 
$\gamma^{\prime \prime}$ to $\tilde \gamma $; 
now $\gamma^{\prime \prime}$ is to be considered as
a function of $\tilde \gamma $.
\be
 \left. \frac{1}{1- \hat \gamma_{\omega}^{(0)} 
\frac{\lambda_1 (\gamma,\ep)}{\gamma} }  
\right|_{\gamma = 0} = \left. 1+ \frac{1}{2\pi i} \oint_{\hat C} 
\frac{d \tilde \gamma}
{\gamma^{\prime \prime} (\tilde \gamma) -\gamma\p} \  
\frac{1}{\tilde \gamma - 
\hat \gamma_{\omega}^{(0) \prime T}  }\ 
\hat \gamma_{\omega}^{(0) \prime T} \ 
\frac{\lambda_1(\gamma^{\prime \prime},\ep) }{1-
\tilde \gamma \lambda_1\p (\gamma^{\prime \prime},\ep)} 
\right|_{\gamma\p = 0} \;.
\ee
We have 
\be \hat \gamma_{\omega}^T = 
\hat \gamma^{\prime \prime} (\tilde \gamma )|_{\tilde \gamma= \hat \gamma_{\omega}^{(0) T}}, \ee
for the solution of the equation (\ref{gammao})
and continue the calculation as
\bea \left. \frac{1}{1- \hat \gamma_{\omega}^{(0)} 
\frac{\lambda_1 (\gamma,\ep)}{\gamma} }  
\right|_{\gamma = 0}  &=& \left.
1+ \frac{1}{ \gamma\p - \hat \gamma_{\omega}^T}
\hat \gamma_{\omega}^T
\frac{1 }{1-
\hat \gamma_{\omega}^{(0) T}\lambda_1\p(\hat \gamma_{\omega}^T,\ep)} 
\right|_{\gamma\p = 0} \nonumber \\
&=& \left.  1+ \sum_{N=1}^{\infty}
\frac{1}{\gamma\p} \hat \gamma_{\omega}^T
\frac{1}{\gamma\p} \hat \gamma_{\omega}^T
...
\frac{1}{\gamma\p} \hat \gamma_{\omega}^T
\frac{1 }{1-
\hat \gamma_{\omega}^{(0) T} \lambda_1\p(\hat \gamma_{\omega}^T,\ep)} 
\right|_{\gamma\p = 0}\;.
\eea 
We write $\frac{1}{\gamma\p} = \int_0^1 \frac{d \alpha}{\alpha}
\alpha^{\gamma\p} $ and use the commutation 
relation (\ref{alphagamma}). 
The $N$th term of the series results in
\be \prod_1^N \int_0^1 \frac{d \alpha_-}{\alpha_i} 
\prod_1^N \gamma_{\omega} (\gamma_{\omega}^{(0)} \alpha_1^{\ep}
\alpha_2^{\ep}...\alpha_i^{\ep} ,\ep) 
A( \gamma_{\o}^{(0)} \alpha_1^{\ep} ...\alpha_N^{\ep}, \ep) 
(\alpha_1 ...\alpha_N)^{\gamma\p} \;.
\ee
We change to the integration variables $\beta_i = \prod_1^i \alpha_1^{\ep}$
and obtain
\be \ep^{-N} \int_0^1 \frac{d\beta_1}{\beta_1} 
\gamma_{\omega} (\gamma_{\omega}^{(0)} \beta_1,\ep) 
\int_0^{\beta_1} \frac{d\beta_2}{\beta_2} 
\gamma_{\omega} (\gamma_{\omega}^{(0)} \beta_2,\ep)...
\int_0^{\beta_{N-1}} \frac{d\beta_N}{\beta_N}
\gamma_{\omega} (\gamma_{\omega}^{(0)} \beta_{N},\ep)
A( \gamma_{\o}^{(0)}\beta_N , \ep)
\beta_N^{\frac{\gamma\p}{\ep}} \nonumber
\ee
\be = \frac{1}{\ep} \int_0^1 
\frac{d\beta_N}{\beta_N}
\gamma_{\omega} (\gamma_{\omega}^{(0)} \beta_N,\ep)
A( \gamma_{\o}^{(0)}\beta_N , \ep)
\beta_N^{\frac{\gamma\p}{\ep}}
\frac{1}{(N-1) !} \left (\frac{1}{\ep} \int_{\beta_N}^1  
\frac{d\beta}{\beta} 
\gamma_{\omega} (\gamma_{\omega}^{(0)} \beta,\ep) \right )^{N-1}.
\ee 
The sum can be done and we obtain at $\gamma\p = 0$
\be
\label{Fresult2}
F= \gamma_{\omega}^{(0)} \left \{
 1+ \frac{1}{\ep}\int_0^1  \frac{d\beta}{\beta}
\gamma_{\omega} (\gamma_{\omega}^{(0)} \beta)
A( \gamma_{\o}^{(0)}\beta, \ep)
 \exp \left(\frac{1}{\ep} \int_{\beta}^1  
\frac{d\beta_1}{\beta_1} 
\gamma_{\omega} (\gamma_{\omega}^{(0)} \beta_1
,\ep )\right) \right \} \;.
\ee
In Appendix D we prove that the asymptotics is obtained 
substituting in the above expression the function
$A(\beta)$ by $A(1)$. 
With this asymptotic result and  restoring also the scale dependence of the dimensionless strong coupling 
$z_{\ep} = (\frac{\kappa^2}{\mu^2})^{\ep} C_{\ep} $ we get
\be
\label{Fas}
F= 
\gamma_{\omega}^{(0)} 
\frac{1 }{1-
 \gamma_{\omega}^{(0)}  
\lambda_1\p( \gamma_{\omega},0)}
\left (\frac{\kappa^2}{\mu^2} \right 
)^{\gamma_{\omega} } 
 \exp \left(\frac{1}{\ep} \int_0^{C_{\ep}}  
\frac{d\beta_1}{\beta_1} 
\gamma_{\omega} (\gamma_{\omega}^{(0)} \beta_1
,\ep )\right) \,  (1 + {\cal O} (\ep) )
\ee
which is equivalent to eq.(\ref{Fresult}),  the renormalization/factorization scale dependence being discussed in the next section.


\subsection{Channels with $n>0$}

Let us now consider the generalisation of the previous calculation for a 
non  vanishing conformal spin $n$. 
Notice first that the singularities of the eigenvalue function
 to the left of ${\cal R}e \gamma = \half$
are located in the vicinity of $\gamma = -\frac{n}{2}$.
For picking up the contributions to the asymptotics
in $\k^2$ it is convenient to move the contour
to this region and change the integration variable to
$\gamma\p = \gamma + \frac{n}{2} $. Then inverse powers
$(\gamma\p)^{-m}$ result in $(\ln \k^2)^m$. 
This shows that the collinear singularity generated
by the partonic impact factor should supply
just a pole in $\gamma\p$ regularised by $\ep$,
i.e. $ (\ep - \gamma\p)^{-1}$. 
For regularisation this pole is kept to the right 
of the contour.

In order to generate this pole  the partonic 
impact factor should be of the form
\be
\label{partonicn}
\Phi_n(\k) = \frac{\bar \alpha_S}{\mu^{2\ep}} 
(\vec \epsilon \vec \k)^n 
(\vec \k^2 )^{-n} \frac{2 }{
|S^{1+2\ep}|}.
\ee 
In analogy to the $n=0$ case we have from 
(\ref{gkkn}, \ref{F})
\bea
\label{F<n}
 F_< (\o, \k, n)\hspace{-0.1cm}&=&\hspace{-0.1cm}\int \frac{d \gamma}{2\pi i}
(\vec \k^2 )^{\gamma -\frac{n}{2} } 
(\vec \epsilon \vec \k )^n  
 \int\frac{d^{2 +2\ep}
\k_0 }{\k_0^2 } \theta(\mu^2_F - \k_0^2)
\frac{1}{ \o - \bar \alpha_S  
e^{-\ep \dd_{\gamma} } \lambda (\gamma,n, \ep) } \nonumber \\
&&\times (\vec \k_0^2)^{ -\gamma - \frac{n}{2} }
(\vec \epsilon^* \vec \k_0)^n \Phi_n (\k_0) 
\\
 &=& \bar \alpha_S (\frac{\mu_F^2}{\mu^2})^{\ep} 
(\vec \k^2)^{-n}(\vec \epsilon \vec \k )^n 
\int \frac{d \gamma\p}{2\pi i}
(\frac{\k^2}{\mu_F^2} )^{\gamma\p  } 
\frac{1}{ \o - \bar \alpha_S 
\frac{1}{\gamma\p}  
\lambda_1 (\gamma\p-\frac{n}{2}, n,\ep) 
    e^{-\ep \dd_{\gamma\p} }  } \nonumber
\frac{1}{\ep - \gamma\p} \;.
\eea
We did the substitution 
$\gamma\p = \gamma + \frac{n}{2}$ and used the function 
$\lambda_1 (\gamma, n, \ep) = \gamma\p 
\lambda (\gamma-\ep, n, \ep) $ defined in eq.(\ref{lambda1n}) 
in  Appendix A. 
Indeed, 
we see that the integral in the second line 
is just the same as in (\ref{F<})
with the only replacement in the
function $\lambda_1$. 
The calculation is therefore completly analogous 
but the change of integration variable reads now
\be
\tilde \gamma = \frac{\gamma\p}
{\lambda_1 (\gamma\p-\frac{n}{2},n, \ep) } 
\ee
which leads to solve, after having picked up the residue in ${\tilde \gamma} = \hat \gamma_{\o}^{(0)} $, the operatorial equation 
\be
 \hat \gamma_{\o}(n,\ep) = \lambda_1 \left(\hat \gamma_{\o}(n,\ep)-\frac{n}{2},n, \ep \right) \hat \gamma_{\o}^{(0)} \; \; 
\ee
 whose solution $\hat \gamma_{\o}$ now expands as power of 
$\hat \gamma_{\o}^{(0)}$ (see the Appendix B).\\
 We finally get, after restoring (as in the $n=0$ case) the $\mu_R$ dependance in the strong coupling 
\be
\label{Fresultn}
 F(\o,\k,n) = 
  \gamma_{\o}(n)
\frac{1}
{ - \gamma_{\o}^2(n) \; \chi_n\p(\gamma_{\o}(n)-\frac{n}{2}) }  
(\frac{\k^2}{\mu_F^2} )^{ \gamma_{\o}(n)}  
\exp \left (\frac{1}{\ep} \int_0^{S_{\ep}} \frac{d\alpha}{\alpha} 
\gamma_{\o} (\gamma_{\o}^{(0)}  (\frac{\mu_R^2}{\mu^2})^\ep  \alpha,n, \ep ) \right )
\ee
where 
\be
\gamma_{\o}(\gamma_{\o}^{(0)},n, \ep ) = 
\gamma_{\o}(\gamma_{\o}^{(0)},n ) + \ep \; \gamma_{\o}^{(\ep)} (\gamma_{\o}^{(0)},n)
+ {\cal O} (\ep^2)
\ee
and the  explicit expressions for $\chi_n$, the BFKL anomalous dimension $\gamma_{\o}( n)$ for conformal spin $n$ and its $\ep$-correction 
$\gamma_{\o}^{(\ep)}( n)$ are given  respectively in Appendix eqs.(\ref{lambda10nchi}), (\ref{BFKLanomdimnexp}), and (\ref{BFKLanomdimepsnexp}).

\section{Analysis of the K-factor}
\setcounter{equation}{0}

\subsection{Extraction of  K-factor }

In order to compare explicitly  the
result (\ref{Fresult}) with 
the result obtained in \cite{CH},  
we define in a  similar way  $R_{\o}$  which absorbs the  ${\cal O} (\ep^0)$ 
factor coming from the exponentiation of the 
$\ep$-correction $\gamma_{\o}^{(\ep)}$ to the BFKL 
anomalous dimension,
\be
\label{Fresultgn}
 F(\o,\k,n) =  \gamma_{\o} (n)\; 
R_{\o}(\gamma_{\o}^{(0)},n)  \; 
(\frac{\k^2}{\mu_F^2} )^{ \gamma_{\o}(n)}\; 
\Gamma \left ( \gamma_{\o}^{(0)} 
(\frac{\mu_F^2}{\mu^2})^\ep , n,\ep \right ).
\ee
This expression singles out the explicit factorization of the collinear singularities (in agreement with \cite{CH}, see also \cite{CFP}), 
appearing in the Laurent series of poles in $\frac{1}{\ep}$ contained in 
the $\overline {MS}$ scheme gluon transition function $\Gamma (\gamma_{\o}^{(0)},n , \ep)$ as a direct consequence of our calculation,
\be
\label{transitionfunctn}
\Gamma (\gamma_{\o}^{(0)},n , \ep) =   
\exp \left (\frac{1}{\ep} \int_0^{S_{\ep}}  
\frac{d\alpha}{\alpha} 
\gamma_{\o}  (\gamma_{\o}^{(0)}  \alpha, n ) \right ).
\ee 
 In eq.(\ref{Fresultgn}) we identify the renormalization and factorization scales ($\mu_R=\mu_F$) in order that 
  $F(\o,\k,n) $ would be  independent of this  arbitrary scale, since 
\be
\Gamma (\gamma_{\o}^{(0)} 
(\frac{\mu_F^2}{\mu^2})^\ep,n, \ep) =   
(\frac{\mu_F^2}{\mu^2})^{\gamma_{\o} (n) } \;
\Gamma (\gamma_{\o}^{(0)},n, \ep) +  
{\cal O} (\ep) \; .
\ee
The BFKL normalization factor $R_{\o}
(\gamma_{\o}^{(0)},n)$ encodes all the dynamics 
coming from the soft singularities resummed by the 
BFKL equation, and which is responsible for the 
singular behaviour of the perturbative QCD Pomeron 
at the saturation value $\gamma_{\o}-\frac{n}{2}=1/2$ 
corresponding to extreme energies as we will see 
further,
\be
\label{Rn}
  R_{\o}(\gamma_{\o}^{(0)},n)=    \frac{1}
{ -\gamma^2_{\o}(n)  \chi_n\p(\gamma_{\o}(n)-\frac{n}{2}) }    \ 
\exp \left ( \int_0^1 \frac{d\alpha}{\alpha} 
\gamma_{\o}^{(\ep)} (\gamma_{\o}^{(0)} \alpha,n) \right ) 
\ee
which can be rewritten explicitly  as 
\be
\label{Rntild}
  R_{\o}(\gamma_{\o}^{(0)},n)=    \frac{1}
{ -\gamma^2_{\o}(n)  \chi_n\p(\gamma_{\o}(n)-\frac{n}{2}) }     \ 
\ee $$
\times \exp \left \{\frac{1}{2} \int_0^{\gamma_{\o}(n)}d\gamma {2 \psi'(1)-\psi'(1+n -\gamma)-\psi'(\gamma) \over
\chi_n(\gamma-\frac{n}{2})}+\chi_n(\gamma-\frac{n}{2})\right\} .
$$
\\

Let us now consider the particular case $n=0$ to make an 
 explicit  comparison with the corresponding K-factor obtained 
by Catani and Hautmann, we have now 
\be
\label{R}
  R_{\o}(\gamma_{\o}^{(0)})=    \frac{1}
{ -  \gamma_{\o}^2 \chi\p(\gamma_{\o}) }    \ 
\exp \left ( \int_0^1 \frac{d\alpha}{\alpha} 
\gamma_{\o}^{(\ep)} (\gamma_{\o}^{(0)} \alpha) \right ) 
\ee
and the gluon transition function
\be
\label{transitionfunct}
\Gamma (\gamma_{\o}^{(0)} , \ep) =   
\exp \left (\frac{1}{\ep} \int_0^{S_{\ep}}  
\frac{d\alpha}{\alpha} 
\gamma_{\o} ( \gamma_{\o}^{(0)}  \alpha ) \right ) \;.
\ee
 For doing this, let us write $R_{\o}$ in a more explicit form by the use of eq.(\ref{gammaeps}) and a change of variable:
\be
\label{Rtild}
  R_{\o}(\gamma_{\o}^{(0)})=    \frac{1}
{ -  \gamma_{\o}^2 \chi\p(\gamma_{\o}) }    \ 
\exp \left \{\frac{1}{2} \int_0^{\gamma_{\o}}d\gamma {2 \psi'(1)-\psi'(1-\gamma)-\psi'(\gamma) \over
\chi(\gamma)}+\chi(\gamma)\right\} .
\ee

\subsection{Comparison of K-factor results}

We  write now the analytical expression of the same quantity, 
obtained in Ref.\cite{CH} by Catani and Hautmann: 
\be
\label{RCH}
R^{CH}(\gamma_{\o}^{(0)})=\left\{{\Gamma(1-\gamma_{\o})\chi(\gamma_{\o})
\over \Gamma(1+\gamma_{\o})[-\gamma_{\o}\chi'(\gamma_{\o})]}\right\}^{1/2}
\exp\left\{\gamma_{\o}\psi(1)+\int_0^{\gamma_{\o}}
d\gamma {\psi'(1)-\psi'(1-\gamma)\over
\chi(\gamma)}\right\} \,.
\ee
To make an explicit comparison with our result, we then use the following identity
\be
\gamma_{\o}\psi(1)+\int_0^{\gamma_{\o}}
d\gamma {\psi'(1)-\psi'(1-\gamma)\over
\chi(\gamma)}= \frac{1}{2} \int_0^{\gamma_{\o}}d\gamma {2 \psi'(1)-\psi'(1-\gamma)-\psi'(\gamma) \over
\chi(\gamma)}+\chi(\gamma)
$$ $$
-\ln \left \{ \gamma_{\o} \chi(\gamma) \frac{\Gamma(1-\gamma_{\o})}{ \Gamma(1+\gamma_{\o})} \right\}^{\frac{1}{2} }
\ee
to  write finally 
\bea
\label{RCH2}
 R^{CH}(\gamma_{\o}^{(0)})&=&\frac{1}{\gamma_{\o} \sqrt{-\chi'(\gamma_{\o})}} \nonumber
\exp \left \{\frac{1}{2} \int_0^{\gamma_{\o}}d\gamma {2 \psi'(1)-\psi'(1-\gamma)-\psi'(\gamma) \over
\chi(\gamma)}+\chi(\gamma)\right\} \,\\
&=&\frac{1}{\gamma_{\o} \sqrt{-\chi \p(\gamma_{\o})}}
\exp \left ( \int_0^1 \frac{d\alpha}{\alpha} 
\gamma_{\o}^{(\ep)} (\gamma_{\o}^{(0)} \alpha) \right ). \,
\eea 
We see that both results agree for the exponentiation of 
the $\ep$-correction $\gamma_{\o}^{(\ep)}$ to the anomalous dimension term, 
but there is a dismatching for the prefactor term. 
We plot these K-factors as a function  of 
$\gamma $ in the physical range $[0,1/2]$:

\begin{figure}[htbp]
\begin{center}
\epsfig{file=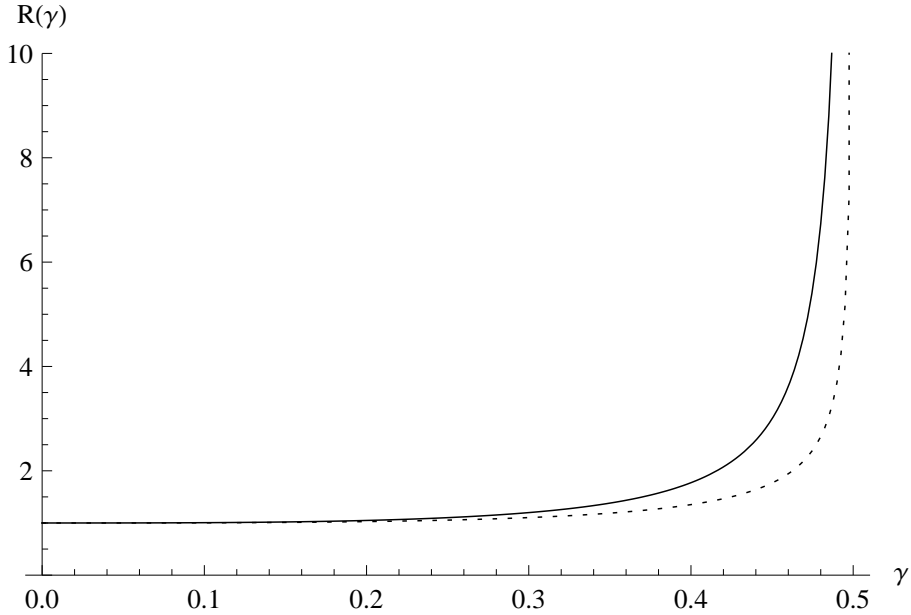,width=12cm}
\end{center}
\caption{K-factor $R(\gamma)$, our result (solid)  
and Catani-Hautmann's result (dashed)}
\end{figure}
We  write the first terms in the coupling expansion for these  factors:
\be
R^{CH}(\gamma_{\o}^{(0)})=1+ \frac{8}{3} \zeta(3) 
\left( \gamma_{\o}^{(0)} \right)^3 - 
\frac{3}{4} \zeta(4) \left( \gamma_{\o}^{(0)}
\right)^4 + \frac{22}{5}  \zeta(5) 
\left( \gamma_{\o}^{(0)} \right)^5  
\ee
$$
+  \left(\frac{209}{9}  \zeta^2(3) -\frac{5}{6} 
\zeta(6) \right) \left( \gamma_{\o}^{(0)} \right)^6 
+ {\cal O} \left( (\gamma_{\o}^{(0)})^7 \right) 
$$
and 
\be
R_{\o}(\gamma_{\o}^{(0)})=1+ \frac{14}{3} \zeta(3) 
\left( \gamma_{\o}^{(0)} \right)^3 - \frac{3}{4} 
\zeta(4) \left( \gamma_{\o}^{(0)} \right)^4 + 
\frac{42}{5}  \zeta(5) 
\left( \gamma_{\o}^{(0)} \right)^5 
\ee $$
+ \left(\frac{419}{9}  \zeta^2(3) -\frac{5}{6} 
\zeta(6) \right)  \left( \gamma_{\o}^{(0)} \right)^6
+ {\cal O} \left( (\gamma_{\o}^{(0)})^7 \right). 
$$ 

With the unintegrated gluon density $ F(\o,\k)  $ obtained in eq.(\ref{Fresult}), 
the $k_T$ factorization of the physical cross-section writes (in Deep Inelastic Scattering for example)
\begin{equation}
\label{ktfac2}
4 Q^2 \sigma_{\o}(Q^2) = \int \frac{d^{2+2\varepsilon} \k }{k^2} \;\;
{\hat \sigma}_{\o}({\k}^{2}/Q^{2}, \bar \alpha_S (Q^2/\mu^2)^{\ep} ;
\ep ) \; F(\o,\k) \; {\tilde f}^{(0)}_{g,\o} (\mu,\ep )
\end{equation}
where the hard cross section (or impact factor) $\hat \sigma$ is obtained in perturbation theory using $Q^2$ as perturbation scale (corresponding  in the case of DIS to the photon virtuality), and ${\tilde f}^{(0)}_{g,\o}$ is the bare gluon distribution function. The factorization of the transition function shown in eq.(\ref{Fresult}) allows to define the renormalized  gluon distribution which now depends on the arbitrary factorization scale $\mu_F$
\begin{equation}
\label{gluonPDF}
{\tilde f}_{g,\o} (\mu_F)= \Gamma (\gamma_{\o}^{(0)} (\frac{\mu_F^2}{\mu^2})^\ep, \ep) \;\; {\tilde f}^{(0)}_{g,\o} (\mu,\ep )\;.
\end{equation}
After having defined the Mellin transform of the  hard cross-section
\begin{equation}
\label{hN}
h_{\o}(\gamma) = \gamma \int^{\infty}_{0} \frac{d
{\k}^{2}}{{\k}^{2}}\left(\frac{\k^{2}}{Q^{2}}\right)^\gamma
\;\; {\hat \sigma}_{\o}({\k^{2}}/Q^{2}, \bar \alpha_S; \ep = 0) \;
\end{equation} 
we finally obtain
\begin{equation}
4Q^{2} \;\sigma_{\o}(Q^{2}) = h_{\o}(\gamma_{\o}(\gamma_{\o}^{(0)}))
\;R_{\o}(\gamma_{\o}^{(0)}) \;(Q^{2}/\mu^{2}_{F})^{\gamma_{\o}} 
\;\; {\tilde f}_{g,\o} (\mu^{2}_{F})\;.
\label{4M2si}
\end{equation} 
Note also  that we can easily interpret the fact that the anomalous dimension appears in the expression 
(\ref{Fresult}) of the  unintegrated gluon density $ F(\o,\k)$, since we can from this Green function define 
the gluon density $ \, G^{(0)} \,$ integrated up to a renormalization scale $\mu_R$: 
\bea
\label{integrG}
{G}_{g g , \, \o}^{(0)}(\bar \alpha_S  , \,\ep)
&=& \int \frac{d^{2+2\ep} \k}{\k^2}
\; F(\o,\k) \; \Theta (\mu_R^2 - \k^2) \;\;
\\ \nonumber
&=& R_{\o}(\gamma_{\o}^{(0)})  \; 
(\frac{\mu_R^2}{\mu_F^2} )^{ \gamma_{\o} }\; \Gamma \left ( \gamma_{\o}^{(0)} (\frac{\mu_F^2}{\mu^2})^\ep , \ep \right ) +  {\cal O} (\ep) \;.
\eea 
If we now identify this renormalization scale with the factorization scale previously defined, $ \, \mu_R^2=\mu_F^2 $, and neglect ${\cal O} (\ep)$ terms, we obtain the following simple expression of the factorized gluon density, equivalent to that defined by Catani and Hautmann (in \cite{CH}, eqs.(3.8) and (3.9) with $ \mu_F^2 $ written $Q^2$) but with $R_{\o}$ instead of $R^{CH}$ :
\begin{equation}
{G}_{g g , \, \o}^{(0)}(\bar \alpha_S  , \,\ep)
=R_{\o}(\gamma_{\o}^{(0)})  \;  
\Gamma \left ( \gamma_{\o}^{(0)} (\frac{\mu_F^2}{\mu^2})^\ep , 
\ep \right ).
\end{equation} 


\subsection{Discussion of the discrepancy}

Let us now summarize how this formulation leads to the factorization of the whole BFKL dynamics into the  K-factor, build 
from two different pieces: the prefactor term and the exponent of the ${\cal O}(\ep)$ part of the BFKL anomalous dimension. The prefactor emerges by picking up the leading twist residue $\gamma_{\o}(\ep)$ of the BFKL green function in $d=2+2\ep$ dimensions and is equivalent in the $\ep$ asymptotics to the one we usually get in exactly two dimensions \cite{CCH,ForshawRoss}. Two ingredients leads to the second term: solving  the leading logarithm BFKL equation in $d=2+2\ep$ transverse dimensions, we notice that $ \bar \alpha_S $ is  appearing with the shift operator $e^{-\ep \dd_{\gamma} }$ in the BFKL Green function. Note that this  feature can be interpreted closely  to the formulation of \cite{ABF}. The action of these involved shift operators on the singularity in $\gamma$ emerging after the convolution of the Green function with the partonic impact factor, leads directly to the exponentiation of the  BFKL anomalous dimension. And since  the   eigenfunctions  $\lambda_1 (\gamma,\ep)$ of the BFKL  kernel gets an $\ep$ correction in $d$ dimensions, it is also the case for the BFKL anomalous dimension $\gamma_{\o}(\ep)$, whose  ${\cal O}(\ep)$ term $\gamma_{\omega}^{(\ep)}$  appears at  ${\cal O}(\ep^0)$ in the  K-factor. 

Focusing the discussion on the gluon density, we see that these two ingredients are in agreement with the renormalisation group requirement going from  $d=2$ to the $d=2+2\ep$ case: in two dimensions, the naive scaling property of the gluon density being  $\left( \frac{\mu_R^2}{\mu^2} \right )^{\gamma_{\omega} }$, must be replaced  in $d=2+2\ep$  transverse dimensions  where the BFKL strong coupling is now  running, see eq.(\ref{alphar}), by 
\be
 G_{gg, \omega} (\mu_R^2, \mu^2, \ep) = 
\bar G_{\omega} (\mu^2, \ep )
\exp \left( \int_{\mu^2}^{\mu_R^2} \frac{d\k^2}{\k^2} 
\gamma_{\omega} (\bar \alpha_S^r(\kappa^2), \ep)
\right) \;.
\ee
The collinear sigularities emerge, if the integration
is extended to $\k= 0$. Indeed
\bea
\int_0^{\mu_R^2}  \frac{d\k^2}{\k^2}
\gamma_{\omega} (\bar \alpha_S^r(\kappa^2), \ep)
= \frac{1}{\ep} \int_0^1 \frac{d\beta}{\beta}
\gamma_{\omega} (\bar \alpha_S \,\beta, \ep)
 +
\int_{\mu^2}^{\mu_R^2} \frac{d\k^2}{\k^2}
\gamma_{\omega} (\bar \alpha_S^r(\kappa^2), \ep) .
\eea 
Thus expanding the anomalous dimension obtained from BFKL in $d$ dimensions as $\gamma_{\o}(\alpha,\ep)=\gamma_{\o}(\alpha) + \ep \, \gamma_{\o}^{(\ep)}(\alpha) + {\cal O} (\ep^2)$, we obtain the bare distribution involving the
singular gluon transition function,
\be
G^{(0)}_{gg, \omega} (\mu_R^2, \mu^2, \ep) = 
\tilde G_{\omega} (\mu_R^2, \mu^2, \ep) \Gamma (\gamma_{\o}^{(0)} , \ep),
\ee
with
\be
\tilde G_{\omega} (\mu_R^2, \mu^2, \ep) = 
\bar G_{\omega} (\mu^2, \ep)
\left( \frac{\mu_R^2}{\mu^2} 
\right )^{\gamma_{\omega} }
\exp \left(\int_0^1 \frac{d\beta}{\beta} 
\gamma_{\omega}^{(\ep)} (\bar \alpha_S\,\beta) \right).
\ee 
The last factor combined with the prefactor coming from the K-factor in 2 dimensions give $R_{\o}$, see eq.(\ref{R}).
\\

We could also observe, that whereas the prefactor 
in the Catani-Hautmann result leads near the 
saturation region (located at the singularity in the twist 2 BFKL anomalous dimension $\gamma_{\o} \simeq 1/2 $ or equivalently  to the square-root singularity in  $ \o \simeq \o_0= 4 \as \ln 2$) to 
a behaviour
\begin{equation}
R^{CH}\simeq \frac{1}{(\o-\o_0)^{\frac{1}{4}}}
\end{equation}
related to a  $(\frac{1}{x})^{\o_0} (\ln(\frac{1}{x}))^{-\frac{3}{4}} $ behaviour in the x-space,
the prefactor in our approach is fully compatible with the square-root branch point singularity typical of the BFKL solution
\begin{equation}
R_{\o} \simeq \frac{1}{(\o-\o_0)^{\frac{1}{2}}}
\end{equation}
 corresponding to the well-known leading behaviour
$(\frac{1}{x})^{\o_0} 
(\ln(\frac{1}{x}))^{-\frac{1}{2}} $.
\\

The discrepancy in the K-factor results is caused
by non-commutative limits, the perturbative 
($\alpha_S \to 0 $) and the regularisation
($\ep \to 0 $) limits. $ F(\o, \k) \ 
\Gamma^{-1} (\gamma_{\o}^{(0)}, \ep) $ has a finite 
limit for $\ep \to 0$. In \cite {CH} this limit has been
calculated resulting in 
$$ \lim_{\ep \to 0}  F(\o, \k) \ 
\Gamma^{-1} (\gamma_{\o}^{(0)}, \ep) = 
\gamma_{\o} \ R^{CH} . $$
On the other hand $\Gamma (\gamma_{\o}^{(0)}, \ep)$
is also the singular factor in the $\ep \to 0 $ asymptotics 
of $ F(\o, \k) $, 
$$ F(\o, \k) = \gamma_{\o} \ R_{\o} (\gamma_{\o}^{(0)})
\Gamma (\gamma_{\o}^{(0)}, \ep)
\left ( 1 + {\cal O} (\ep) \right ). $$
As we have seen in the calculation, the prefactor in the 
asymptotics does not coincide with the results of the
limit, $ R_{\o} \not = R_{\o}^{CH}$. 
  
For illustration consider instead of the actual BFKL 
equation 
the simplified case without extra $\ep$ dependence in
the eigenvalue function $\frac{\lambda_1(\gamma)}{\gamma}
= \chi (\gamma)$ and substitute $\lambda_1 (\gamma)$
by simple expressions. The  results (\ref{R}, \ref{RCH2})
reduce to
$$ 
\gamma_{\o} \ R_{\o} (\gamma_{\o}^{(0)}) =  
\frac{1}{- \gamma_{\o} \chi\p (\gamma_{\o})},
\ \ \ \ 
\gamma_{\o} \ R_{\o}^{CH} (\gamma_{\o}^{(0)}) = 
\left (
\frac{ 1}{ - \chi\p (\gamma_{\o})}
\right )^{\half},
$$
where as above  $\gamma_{\o}$ is the solution of 
$ 1 = \gamma_{\o}^{(0)} \chi (\gamma) $. 

In the particular example 
$\lambda_1 (\gamma) = 1 + a \gamma^3$
and small coefficient $a \ll 1$ we write the first terms 
in the perturbative expansion of $F $ 
$$ F = \gamma_{\o}^{(0)} \left \{
1 + \gamma_{\o}^{(0)} (\frac{1}{\ep} + a \ep^2) +
\gamma_{\o}^{(0) 2} (\frac{1}{ 2 \ep^2} + a \frac{9}{2} 
\ep )
+ \gamma_{\o}^{(0) 3} (\frac{1}{ 6 \ep^3} + a 6) + ...
\right \}
$$
Factorise the singular terms related to the expansion of
$ e^{\frac{1}{\ep} \gamma_{\o}^{(0)} } $,
$$ \gamma_{\o}^{(0)}
( 1 + \gamma_{\o}^{(0)} a \ep^2 + 
\gamma_{\o}^{(0) 2} a \frac{7}{2} \ep + 
2 \gamma_{\o}^{(0) 3} a + ...)
( 1 + \gamma_{\o}^{(0) } \frac{1}{\ep} +
\gamma_{\o}^{(0) 2} \frac{1}{2\ep^2} +
\gamma_{\o}^{(0) 3} \frac{1}{ 6 \ep^3} + ...)
$$
After separating the singularities in this way
the $\ep^0$ terms in the factor in front of the 
singular one approximate
$\gamma_{\o} \ R_{\o}^{CH} $.  

Our calculation is focussed on the asymptotics. 
In this case the terms with positive powers of $\ep$
are to be neglected from the start. This leads to
a difference in separating the singularities order by order
$$ \gamma_{\o}^{(0)}
( 1 +  
\gamma_{\o}^{(0) 3}\, 6\, a   + ...)
( 1 + \gamma_{\o}^{(0) } \frac{1}{\ep} +
\gamma_{\o}^{(0) 2} \frac{1}{2\ep^2}
+ \gamma_{\o}^{(0) 3} \frac{1}{ 6 \ep^3} + ...).
$$

The relation of the discrepancy to the positive power
$\ep$ terms is confirmed by the example
$\lambda_1 (\gamma) = 1 + a \gamma$ ( $a$ not small).
No terms with positive powers of $\ep$ appear here.
The sum in $F$ can be done here. Thus 
the K-factor expressions are easily checked and we obtain
$R_{\o} = R_{\o}^{CH}$ for this particular case.

The extraction of the collinear singularities along the
scheme by Catani and Hautmann has been reanalysed
by Ciafaloni and Colferai \cite{CC} confirming the 
previous result with the  preexponential 
factor $\sim (\chi\p (\gamma_{\o})^{-\half}$. 

The $Q_0$ regularisation is used there. 
It  means to introduce
in the convolution with the parton impact factor
besides of the factorisation scale $\mu_F$ an
infrared cut-off $Q_0^2 = \mu_F^2 e^{-T}$ 
for large $T$. 
In that  paper first the
Green function of the BFKL equation is derived.
This step is analogous to our approach, 
but there   the solution is derived by iteration 
in steps of size $\ep$ and the result 
is written in Mellin transformation with respect to 
$\k, \k_0$. 
The iteration starts from a Born term involving the 
cut-off $Q_0$ (eq.(2.9) in \cite{CC}), 
$$ f^{(0)}_{Q_0} = \gamma_{\o}^{(0) }
\frac{e^{\gamma T}}{\gamma}, T= \ln \frac{\mu^2}{Q_0^2}. $$

The calculation of \cite{CC} can be easily adapted to our 
scheme allowing to see the discrepancy appearing 
from another side.
We should replace this Born term by the one calculated from the
parton impact factor (\ref{partonicIF}) with the result
$$ f^{(0)}_{Q_0} = \gamma_{\o}^{(0) } 
\frac{1}{\ep - \gamma} =
\int_0^{\infty} dT e^{-    \ep T} 
\gamma_{\o}^{(0) }
e^{\gamma T}. $$
Notice that now $T$ is an integration variable with no 
relation to $Q_0$. The dominant contribution arises from
large values $T\sim \frac{1}{\ep}$.

This replacement leads to the following modification
of the solution (eq (2.29) in \cite{CC}, where
the $\ep$ dependence in the eigenvalue function is 
suppressed now),
$$ \tilde F_{\ep} (t) = \int_0^{\infty} dT e^{-\ep T}
\int \frac{d\gamma }{2 \pi i} \int^{\gamma} d \gamma\p 
\frac{\exp (\gamma t + \frac{1}{\ep} 
\int_{\gamma\p}^{\gamma} L_0 (z) dz - \gamma\p T ) }
{ \ep \sqrt{ \chi(\gamma)} \sqrt{\chi (\gamma\p)} }
\ \frac{L_0 (\gamma\p) - \ep T}
{1 - e^{-L_0 (\gamma\p) + \ep T}}. $$
The notation $L_0 (\gamma) = 
\ln (\gamma_{\o}^{(0)} \chi(\gamma)) $ is used here.
The estimate of the asymptotics at $t = \ln {\k^2}{\mu^2}
\gg 1$ and at $T \sim \frac{1}{\ep} \gg 1$ is done by
applying the saddle-point approximation twice. The
saddle-point equations are similar,
$$ t + \frac{1}{\ep} L_0(\gamma) =0, \ \ \ 
T- \frac{1}{\ep} L_0 (\gamma) = 0 
$$
with the solutions
$$ \gamma : \ \ \bar \gamma_t = 
\gamma_{\o} (\gamma_{\o}^{(0)} e^{\ep t}), \ \ \ 
\gamma\p : \ \ \bar \gamma_T =   
\gamma_{\o} (\gamma_{\o}^{(0)} e^{-\ep T}).
$$
$\gamma_{\o} (\gamma_{\o}^{(0)} )$ denotes the solution of
$1 =  \gamma_{\o}^{(0)} \chi (\gamma)$. 
The integral over the fluctuation results in two factors
$ (\frac{\ep \chi(\bar \gamma)}
{ -\chi\p (\bar \gamma)} )^{\half}$, The estimate for 
$\tilde F$ is therefore
$$ \tilde F_{\o} (t) =
\int_0^{\infty} dT e^{- T (\ep- \bar \gamma_T) }
\frac{\exp (\bar \gamma_t t + \frac{1}{\ep} 
\int_{\bar \gamma_T}^{\bar \gamma_t} L_0 (z) dz 
 ) }
{  \sqrt{ \chi\p(\bar \gamma_t)} \sqrt{\chi\p (\bar \gamma_T)} }
\ \frac{L_0 (\bar \gamma_T) - \ep T}
{1 - e^{-L_0 (\bar \gamma_T) + \ep T}} $$

The second saddle point does not tend to zero in the
regularisation limit and both fluctuation factors remain.

In the analysis of \cite{CC} with $Q_0$ 
the second fluctuation factor turns to 1, if the limit
$Q_0 \to 0$ is taken before the $\ep$ regularisation limit
because then the corresponding saddle point tends to zero.
To match with our asymptotic factorisation the $
\ep$ asymptotics is to be considered, where $\ep =0$ in 
all regular terms.
At vanishing $\ep$ the saddle point does not tend to zero
and no fluctuation factor turns to 1.

\section{ Structure functions and
exclusive electroproduction}
\setcounter{equation}{0}

\subsection{The longitudinal structure function $F_L$}

As an application of the previous discussion, we can  write the factorized expression of the longitudinal structure function $F_L$ in the Mellin space, in agreement with \cite{CH}

\begin{equation}
\label{FL}
F_L^{\o} (Q^{2})=C_{L , \, {\o}}^g(\as , Q^{2}/\mu^{2}_{F})  \;\; f^g_{\o}(\mu^{2}_{F}) \;,
\end{equation}
where $f^g_{\o}(\mu_{F}^{2})$ is the ${\o}-$moment of the renormalized gluon distribution function and we have defined  the gluonic (improved) coefficient function 

\begin{equation}
C_{L , \, {\o}}^g(\as , Q^{2}/\mu^{2}_{F}) =
h_{L , \, {\o}} \left( \gamma_{{\o}} \right) \;
R_{{\o}} \;(Q^{2}/\mu^{2}_{F})^{\gamma_{{\o}}} 
\label{CLNg}
\end{equation}
with the Mellin transform of the corresponding hard cross-section 

\begin{equation}
\label{hL}
 \, h_{L, \, {\o} } ( \gamma_{\o}) = {  \bar \alpha_S \over { 2 \, \pi}} \, N_f \, T_R \,
{ { 4 \, ( 1 - \gamma_{\o}) } \over { 3 - 2  \gamma_{\o} }} \, { {
\Gamma^3 ( 1 - \gamma_{\o})} \over {
\Gamma ( 2 - 2 \gamma_{\o}) } }  { { \Gamma^3 ( 1 + \gamma_{\o}) } \over {
 \Gamma ( 2 + 2 \gamma_{\o}) } } \;.
\;\;
\end{equation}
Then we  easily obtain from (\ref{R}) and 
(\ref{BFKLanomdim}) the all order  
perturbative expansion in power of  
$\gamma_{\o}^{(0)}$ of this coefficient 
function for the simpler case $\mu^{2}_{F}=Q^2$
\bea
\label{CMSL}
&~&{C_{L , \, N}^{g}}(\as , {Q^2/\mu_F^2} = 1 )
= \frac{\as}{2\pi} T_R N_f { 4 \over 3}
\left\{ 1 - \frac{1}{3} \gamma_{\o}^{(0)} +
\left[\frac{34}{9}-  \, \zeta (2)    \right]
\left(\gamma_{\o}^{(0)} \right)^2 + \left[ - \frac{40}{27} \right. \right.
\nonumber
\\
&~&+
\left.
{1 \over 3} \, \zeta (2) +
{14 \over 3} \, \zeta (3)    \right] \,
 \left(\gamma_{\o}^{(0)} \right)^3 +
\left[  \frac{1216}{81}-
{34 \over 9} \, \zeta (2) -
{20 \over 9} \, \zeta (3)  - 6 \, \zeta (4)  \right] \,
 \left(\gamma_{\o}^{(0)} \right)^4
\nonumber\\
&~&\left. + \, {\cal O}\left(\left(\gamma_{\o}^{(0)} \right)^5\right) \right\}
\\
&~&\simeq \frac{\as}{2\pi} T_R N_f \frac{4}{3} \left\{ 1 - 0.33 \gamma_{\o}^{(0)} +
2.13 \left(\gamma_{\o}^{(0)} \right)^2 +
4.68 \left(\gamma_{\o}^{(0)} \right)^3 
-0.37 \left(\gamma_{\o}^{(0)} \right)^4 \right. \nonumber \\
&~&+ \left.
{\cal O}\left(\left(\gamma_{\o}^{(0)} \right)^5\right) \right\} \;\;, \nonumber
\eea
which has to be compared with eq.(5.24) of 
Catani-Hautmann's paper \cite{CH}. 
Indeed, the results start to deviate 
 at the fourth loop in the  perturbative
expansion. 
The calculation made by 
Moch, Vermaseren and Vogt \cite{Moch}
on the same coefficient function derived by complete 
loop calculations in pure collinear factorization 
scheme extends to three loops and is, therefore, 
still not sufficient to discriminate the small $x$
resummation results.\\

\begin{figure}[htbp]
\begin{center}
\epsfig{file=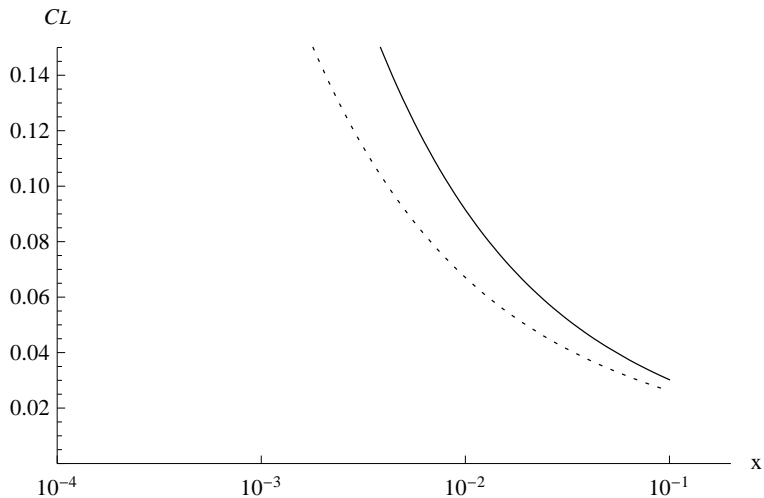,width=10cm}
\end{center}
\caption{$C_{L , \, N}^{g}$,\; $Q^2 = \mu^{2}_{F}$}
\label{curve0}
\end{figure}

Going from $(\o, \gamma)$ to the $(x,Q^2)$ space, we display in   Fig.\ref{curve0} the longitudinal coefficient functions, the dotted curve corresponding to the Catani-Hautmann's result and the solid one to our approach. We considered for that the case $N_f=4$, $\log Q^2/\Lambda^2=6$ in the strong coupling and $\mu^{2}_{F}=Q^2$. We show this plot also for comparison with \cite{WT}, in particular with the  Fig.6 which gives the corresponding result for $C_L$. Note that we restrict the comparison to the small-x ($x \le 10^{-1}$) region to avoid the discussion of modelling the Born $\delta(1-x)$ term. The analysis of \cite{WT} shows the impact of the running coupling (see \cite{T}) and of the large corrections coming from  NLL BFKL resummation which essentially tame the small-x growth: the authors  consider the NLL BFKL equation \cite{NLL} with leading order running coupling and an estimation of the longitudinal NLL impact factor.  The NLL BFKL extension of our approach goes beyond the scope of our present study; it would be needed to confirm whether those corrections imply a weaker small-x growth like in the result of \cite{WT}.
\\

\begin{figure}[htbp]
\begin{center}
\epsfig{file=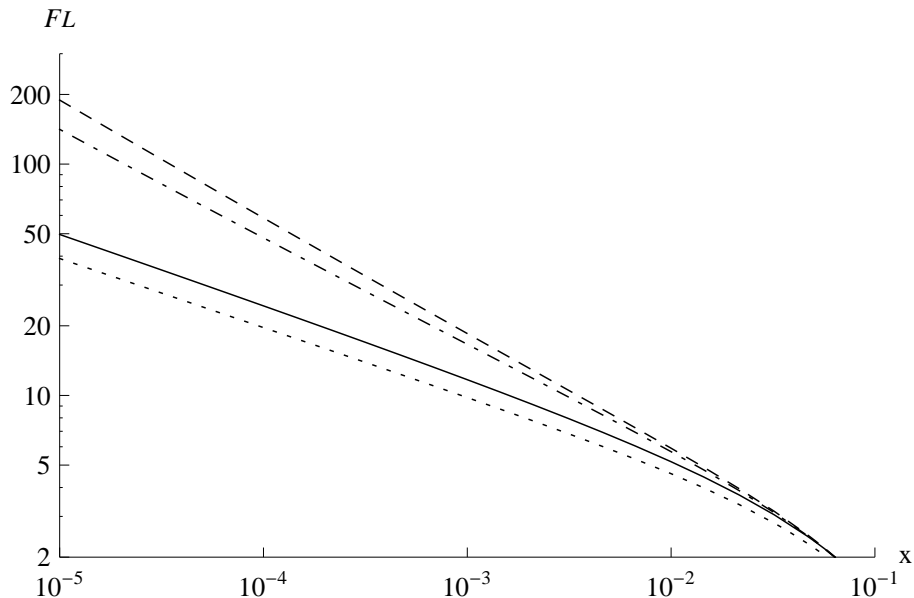,width=12cm}
\end{center}
\caption{$F_L$, $Q^2 = 30 GeV^2$}
\label{curve1}
\end{figure}

\begin{figure}[htbp]
\begin{center}
\epsfig{file=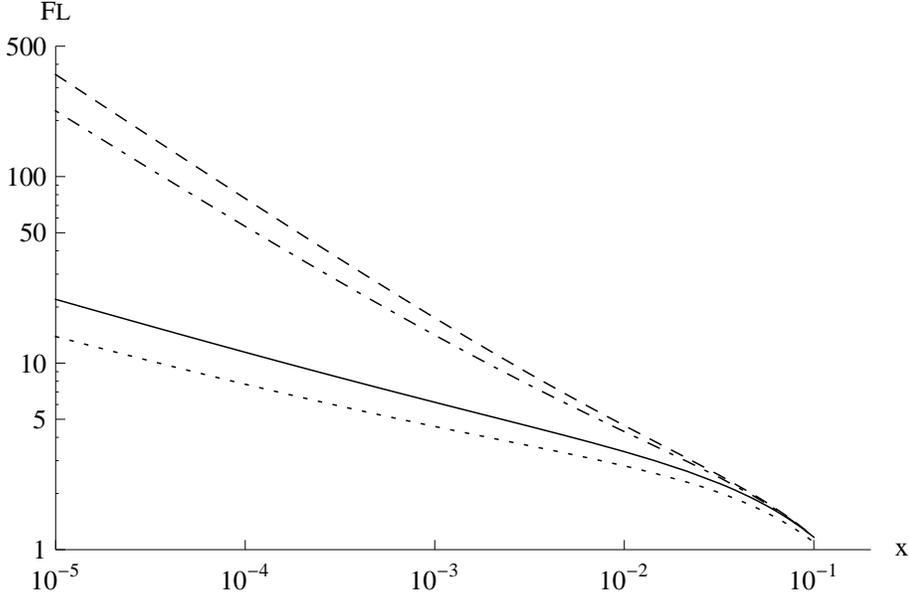,width=12cm}
\end{center}
\caption{$F_L$, $Q^2 = 10 GeV^2$}
\label{curve2}
\end{figure}

In order to estimate the order of the discrepancy of our approach with the one of Catani-Hautmann for the longitudinal structure function, we consider  in the following $F_L (x, Q^{2})$ 
as a function of $x$ for 
different values of the hard scale $Q^2$. 
We simply used for the convolution with the soft part the 
following parametrization of the gluon 
PDF $x g(x, Q_0^2)$ at $Q_0^2=30 GeV^2$ (cf. \cite{Moch})
\begin{equation}
\label{gPDF}
x g(x, Q_0^2)=1.6\; x^{-0.3} \;(1-x)^{4.5} \;(1-0.6 \;x^{0.3})
 \end{equation}
and we have made it evolve through the DGLAP equation in the 
small $x$ limit at one loop accuracy to obtain its $Q^2$ 
 dependence. We then obtain the following curves Fig.\ref{curve1} and Fig.\ref{curve2}.

The solid curve corresponds to the Born term in 
the high energy expansion, equivalently to the 
leading order accuracy. The dotted curve contains 
the Born term and the negative 
(see eq.(\ref{CMSL})) next-to-leading order 
corrections, explaining the trend. 
Then we do the expansion of the small $x$ resummation
result in $\bar \alpha_S \ln x$ up to 12th order
which is quite sufficient because of good convergence.
This allows to show the discrepancy
for these phenomenological predictions between the 
two different K-factor expressions: 
the dashed curve corresponds to our result and 
the dashed-dotted one to the same analysis 
with the  Catani-Hautmann result. 
The convergence is very convincing and it is 
increased with the value of the hard scale.
All the curves are plotted  with the factorization 
scale $\mu^{2}_{F}=Q^2$.

\subsection{Exclusive electroproduction}

\begin{figure}[htbp]
\begin{center}
\epsfig{file=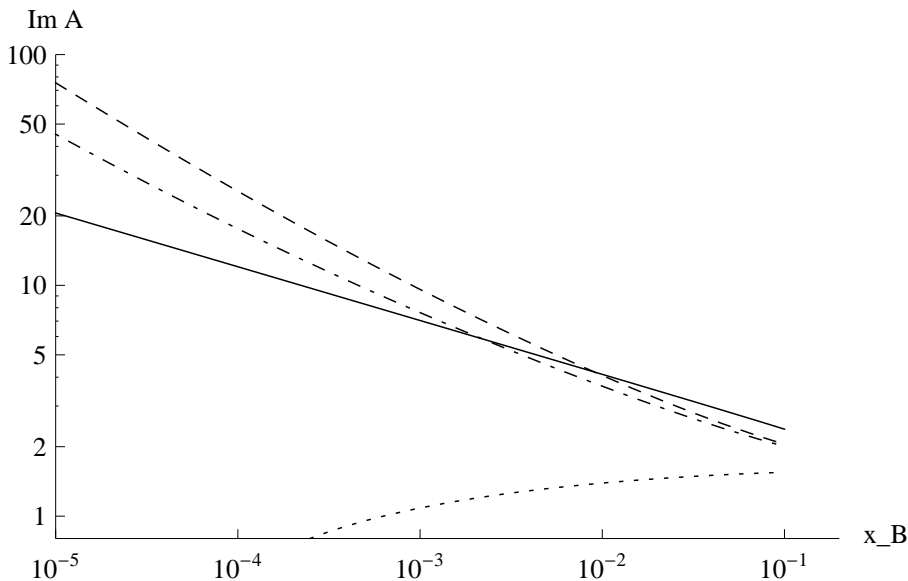,width=12cm}
\end{center}
\caption{ VM electroproduction,
$Q^2 = 30 GeV^2$}
\label{curve4}
\end{figure}

\begin{figure}[htbp]
\begin{center}
\epsfig{file=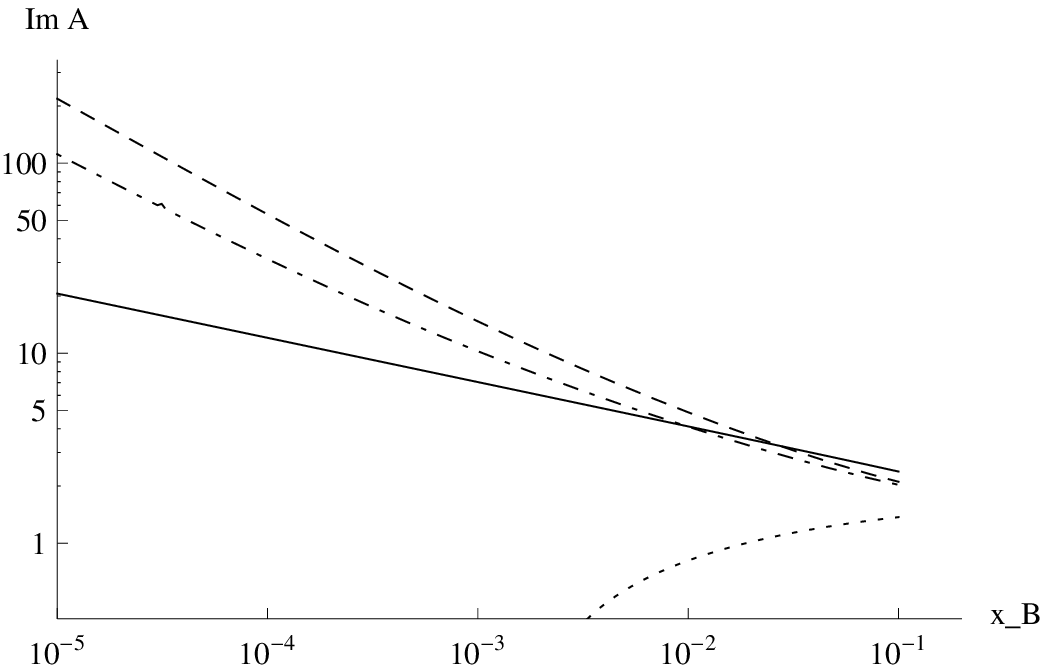,width=12cm}
\end{center}
\caption{VM electroproduction,
$Q^2 = 10 GeV^2$}
\label{curve5}
\end{figure}

We now turn to the application for exlusive vector 
meson (VM) electroproduction in 
Deeply Virtual Compton Scattering (DVCS) 
following the line of  \cite{Ivanov:2007je}: 
considering the gluon dominance in the Regge limit 
of the scattering, the amplitude reads 
\be
\label{forcalc}
{\cal I}m A^g \simeq {H^g(\xi,\xi)}
+ \int\limits^1_{\xi}\frac{d x}{x}{H^g(x,\xi)}
 \, \sum\limits_{n=1}C^g_{VM,n}\frac{\bar \alpha_s^n}{(n-1)!}\log^{n-1}\frac{x}{\xi} \, ,
\ee
where ${H^g(\xi,\xi)}$ is the Born contribution of 
the gluon GPD. The $C^g_{VM,n}$ are polynomials 
of  $\log\frac{Q^2}{\mu_F^2}$ obtained as in the 
previous example in the perturbative expansion of 
the coefficient function in power of 
$\gamma_{\o}^{(0)}$
\bea
\label{CLVMg}
C_{VM , \, {\o}}^g(\as , Q^{2}/\mu^{2}_{F})  \hspace{-0.2cm}&=& \hspace{-0.2cm}
h_{VM , \, {\o}} \left( \gamma_{{\o}}  \right) \;
R_{{\o}} \;(Q^{2}/\mu^{2}_{F})^{\gamma_{{\o}}} \nonumber \\
&=&  \hspace{-0.2cm}\sum\limits_{n=0} C^g_{VM,n} (\gamma_{\o}^{(0)})^n \;\;,
\eea
with the following expression of the Mellin 
transform $h_{VM , \, {\o}} $ of the hard 
cross-section: we define for that the properly 
normalized impact factor $\gamma^* \to VM $ written 
as a convolution of  the hard scattering 
amplitude  with the leading twist non-perturbative 
Distribution Amplitude (DA) \cite{DA}, where both 
photon and vector meson are longitudinally polarized. 

\begin{equation}
\label{liVM}
h_{VM}(k_t^2)=\frac{\int\limits^1_0 dz \,
\frac{Q^2}{k_t^{\,\, 2}+z (1-z) Q^2}\phi_{VM} (z)
}{\int\limits^1_0 dz \,\frac{\phi_{VM} (z)}{z(1-z)} }\, ,
\end{equation} 
and its Mellin transform reads for an asymptotic vector meson 
DA $\phi_{VM} (z)= 6 z (1-z)$,
\bea
\label{galiVM}
\nonumber
 h_{VM , \, {\o}}(\gamma_{\o}) \hspace{-0.2cm}&=&\hspace{-0.2cm}\gamma_{\o} \int\limits^\infty_0\frac{d  k_t^{\,\, 2}}{k_t^{\,\, 2}}
\left(\frac{k_t^{\,\, 2}}{Q^2}\right)^{\gamma_{\o}} h_V(k_t^2)
\\
&=&\hspace{-0.2cm} \frac{\Gamma^3(1+\gamma_{\o})\Gamma(1-\gamma_{\o})}{\Gamma(2+2\gamma_{\o})}\, .
\eea
We replace in the high energy term the gluon GPD by its forward 
limit  {$H^g(x,\xi)\to x g(x)$}. On the contrary to the previous  
study for $F_L$, we keep this expression for the soft part without 
doing any $Q^2$-evolution. We replace the Born term by a very simple 
model $H^g(\xi,\xi)=1.2  \xi g(\xi)$. We obtain the following curves Fig.\ref{curve4} and Fig.\ref{curve5}, 
in the same spirit as  previously by doing the high energy expansion.


The solid curve corresponds to the Born term, 
the dotted curve to   the Born term with the 
NLO corrections. 
Note that the numerical value of the $C^g_{VM,1} $ 
coefficient is much larger than in the $F_L$ case, 
giving these stronger and very negative NLO 
corrections.  This appears as an instability
in the perturbative prediction 
for this process. Here the small $x$ resummation is
really needed to obtain a reasonable and stable
prediction \cite{Ivanov:2007je}.  
After twelve iterations we get the dashed curve 
corresponding to our result and the dashed-dotted 
one corresponding to the same analysis 
with the  Catani-Hautmann K-factor. 
Also here  the convergence of the series 
expansion after twelve iterations is very good.  
Although the curves are only plotted  here with 
the factorization scale $\mu^{2}_{F}=Q^2$, 
we observe that the factorization scale dependence 
is reduced when taking into account the 
high energy resummations compared to the Born or 
even NLO case. We also note that the sensitivity 
of this choice is equivalent for both K-factor 
expressions.

\section{Summary}

The resummation of small $x$ corrections to hard scattering amplitudes
calculated by the collinear factorization method has been 
reconsidered. The contributions enhanced by $\ln x$ are resummed by 
BFKL equation. The collinear singularities appear due to the 
convolution of the BFKL Green function  with a parton 
impact factor. The factorization of these singularities 
is  demonstrated in accordance with the scheme chosen in the collinear
calculation. Besides of the resummed gluon anomalous dimension
a correction factor is derived that amounts into the improved
coefficient function. In this way we follow the known resummation 
scheme. However, our result for the K-factor differs from the one
given in \cite{CH} and used in previous applications. 

We have presented our arguments in all details in order to explain the
origin of the difference related to the non-commutativity of 
the  perturbative and the regularisation limits.
The basic elements are the BFKL Green function in
$2+2\ep$ transverse dimensions, its convolution with
the parton impact factor and the factorisation of the
resulting collinear singularities in the small $\ep$
asymptotics. For comparison we have applied the same 
asymptotic analysis to the iterative solution given in
\cite {CH} and obtained the same result.
Our result has the square-root
singularity in the continued angular momentum $\o$ typical for 
BFKL partial-wave amplitudes. 

The significance of the difference in the K-factor has been illustrated
in applications to structure function and to exclusive electroproduction 
of vector mesons.  Hard exclusive processes as a new kind of applications of
the small $x$ resummation has ben pointed out in \cite{Ivanov:2007je}.
The formulation presented here extends to channels corresponding 
to non-vanishing conformal spin $n$ in the BFKL solution. These appear in
polarized structure functions and contribute to hard exclusive amplitudes.

\section*{Acknowledgments}

We are grateful to D. Yu. Ivanov for useful 
discussions and correspondence. 
This work is  supported by   DFG 
(contract KI-623/4).


\appendix
\section*{Appendices}

\section{Calculation of $\lambda_1(\gamma,n, \ep) $}
\setcounter{equation}{0}

\subsection{Case $n=0$}

We now study the action  of the full kernel in $2 + 2 \ep$ transverse dimensions on the pseudo eigenfunctions $(\vec \k^2)^{\gamma -1}$ 

\be \hat K \cdot (\vec \k^2)^{\gamma -1} =\frac{1}{\pi} \int  \frac{d^{2 + 2\ep}\k\p }{(2 \pi)^{ 2\ep}} \ \left \{ \frac{1}{ (\vec \k - \k\p )^2 } \ 
-\frac{1}{2} \alpha_G (\vec \k^2) \delta^{(2 + 2\ep)} (\vec \k - \vec \k\p) \right \} \ \ 
((\vec \k\p )^2)^{\gamma -1} \;.
\ee 
We need first to evaluate the gluon trajectory function 
\bea
\label{alphaG}
\alpha_G (\k)&=&\int \frac{ d^{2 + 2\ep} \k\p \ \vec \k^2}{
(\vec \k\p )^2 \ (\vec \k - \vec \k\p )^2 } \\
&=&
\pi^{1+\ep} (\vec \k^2 )^{\ep} \frac{ \Gamma (1-\ep) \Gamma^2(\ep) }{\Gamma
(2\ep) }. \nonumber
\eea 
then we calculate the action of the bare kernel 
\bea
\int \frac{ d^{2 + 2\ep} \k\p \ }
{ (\vec \k - \vec \k\p )^2 } \ ((\vec \k\p )^2)^{\gamma -1} 
= \int \frac{ d^{2 + 2\ep} \k\p}{  \Gamma ( 1-\gamma) }
\int_0^{\infty} d\lambda_1 d\lambda_2 \ \lambda_1^{-\gamma}
e^{ - \lambda_1 (\vec \k\p )^2 - 
\lambda_2 (\vec \k - \vec \k\p )^2 }\\
= \pi^{1+\ep} (\vec \k^2 )^{\gamma -1 + \ep} 
\frac{1}{\Gamma (\ep) } B(\ep, 1-\gamma -\ep) \ B(\ep, \gamma + \ep) \;. \nonumber
\eea
With these results we compose the action of the full kernel
\be
\label{kernb}
 \hat K \cdot (\vec \k^2)^{\gamma -1} = 
\lambda (\gamma, \ep) (\vec \k^2 )^{\gamma -1 + \ep}.  
\ee 
We have defined this quasi-eigenvalue function:
\be
\label{lambdagam}
\lambda (\gamma, \ep) = \frac{1 }{(4 \pi)^{\ep}}  
 \left [ b(\gamma, \ep) - \half b(0, \ep) \right ] 
\ee 
with 
\be
\label{bgamma} 
 b(\gamma, \ep) = \frac{1}{\Gamma (\ep) } 
B(\ep, 1-\gamma -\ep) \ B(\ep, \gamma + \ep).
\ee  
Since in our calculation we encounter the shifted quantity
$ \lambda (\gamma-\ep, \ep)$, we define  the function
\bea
\label{lambda1first}
\lambda_1 (\gamma, \ep)&=& \gamma \; \lambda (\gamma-\ep, \ep)
\\ \nonumber
&=&\frac{1 }{(4 \pi)^{\ep}} \Gamma (1+ \ep)  
\left[ \frac {\Gamma (1 - \gamma )}{\Gamma (1-\gamma + \ep)} 
\frac {\Gamma (1+ \gamma )}{\Gamma (1+ \gamma +  \ep)} \right.
\\ \nonumber
 &+&\left.  \frac{\gamma}{\ep} \left ( \frac {\Gamma (1 - \gamma )
}{\Gamma(1-\gamma + \ep)} 
\frac {\Gamma (1+ \gamma )}{\Gamma (1+ \gamma +  \ep)}
 -
\frac {\Gamma (1  + \ep) 
\Gamma (1  - \ep)}{\Gamma (1 + 2\ep)} \right ) \right] \;.
\eea 
and give its expression in the $\overline {MS}$ renormalization scheme
\bea
\label{lambda1}
\lambda_1 (\gamma, \ep)= \gamma \; e^{\ep \psi(1)}
\frac{ \Gamma (1+ \ep)}{\ep}  \left ( \frac {\Gamma (1 - \gamma ) \; \Gamma ( \gamma )
}{\Gamma(1-\gamma + \ep) \;\Gamma ( \gamma +  \ep)}  -
\frac {\Gamma (1  + \ep) 
\Gamma (1  - \ep)}{\Gamma (1 + 2\ep)} \right ) \;.
\eea 
We expand this function in $\ep$ up to the first power (higher orders are suppressed in our approach) and get
\be
\label{lambda1ep}
\lambda_1 (\gamma, \ep) = 
\lambda_1 (\gamma,0) + \ep   \lambda_1^{(\ep)} 
(\gamma) + {\cal O} (\ep^2)
\ee 
where the constant term is 
\be
\label{lambda10}
\lambda_1 (\gamma,0) = 1 + \gamma \left(2 \psi (1) - \psi(1-\gamma) - \psi (1+\gamma) \right)
\ee 
and the contribution proportional to the firstpower of $\ep$ is
\be
\label{lambda1eps}
\lambda_1^{(\ep)} (\gamma) = 
\frac{\gamma}{2}\left (2 \psi\p(1)  - 
\psi\p(1-\gamma) - \psi\p (\gamma)  +  
\chi^2 (\gamma)\right)
\ee
where $\chi(\gamma)$ is the   well known one-loop BFKL eigenvalue function also recovered in 
\bea
\label{chi}
\chi (\gamma) &=&  \frac{1}{\gamma} \ \lambda_1 (\gamma, 0)
 \nonumber \\
&=&2 \psi (1) - \psi(1-\gamma) - \psi (\gamma) \;.
\eea
Also the expansions in powers of $\gamma$ of the previous expressions (\ref{lambda10}) and (\ref{lambda1eps}) reads:
\be
\label{lambda10exp}
 \lambda_1 (\gamma, 0) = 1 + \sum_{k=1}^{\infty} 2 \zeta (2 k +1)
\gamma^{2k +1} 
\ee
and similarly for  $\lambda_1^{(\ep)} $ :
\be
\label{lambda1epsexp}
\lambda_1^{(\ep)} (\gamma) = \sum_{k=1}^{\infty} 2 \zeta (2 k + 1 ) \gamma^{2 k}
 + \sum_{k=1}^{\infty} \left[ - \left(2 k + 1 \right) \zeta (2 k + 2 ) + 2 \sum_{p=1}^{k-1} \zeta (2 p + 1 ) \zeta (2 (k-p) + 1 ) \right] \gamma^{2 k + 1} .
\ee

\subsection{Case $n> 0$}

We calculate now the action of the full kernel on an ansatz for the general conformal spin $n$ pseudo 
eigenfunctions $ ((\vec \k )^2)^{\gamma -1 - \frac{n}{2} } (\vec b . \vec \k)^n $, in order to obtain an analytical expression of the first order $\ep$ correction to the corresponding one-loop BFKL eigenvalue function $\chi_n$. The trajectory of the gluon is not changed, then we need to compute the action of the real (production) part of the kernel on these pseudo eigenfunctions

\bea
&& \int  \frac{ d^{2 + 2\ep} \k\p \ }
{ (\vec \k - \vec \k\p )^2 } \ ((\vec \k\p )^2)^{\gamma -1 - \frac{n}{2}} (\vec b . \vec \k\p)^n \nonumber \\ 
&=& \int  d^{2 + 2\ep} \k\p  \Gamma^{-1} ( 1 + \frac{n}{2} -\gamma) 
\int_0^{\infty} d\lambda_1 d\lambda_2 \ \lambda_1^{\frac{n}{2}-\gamma}
\exp ( - \lambda_1 (\vec \k\p )^2 - 
\lambda_2 (\vec \k - \vec \k\p )^2 )  \ (\vec b . \vec \k\p)^n \nonumber \\
&=& \sum_{i=0}^n \binom{n}{i} \pi^{ \frac{1}{2}+\ep} \frac{[1+(-1)^i]}{2} \frac{\Gamma (\frac{i+1}{2}) \Gamma (1+\frac{n-i}{2} -\gamma - \ep)\Gamma (\gamma + \ep + \frac{n-i}{2})\Gamma (\ep +  \frac{i}{2})  }{\Gamma (1 + \frac{n}{2} -\gamma) \Gamma (\gamma + 2 \ep +\frac{n}{2}) } b^i \nonumber \\
&\times& ((\vec \k )^2)^{\gamma +\ep -1 - \frac{n-i}{2}}(\vec b . \vec \k)^{n-i} \;.
\eea
We now keep only the $i=0$ term corresponding to the  conformal spin $n$ representation  of the pseudo-eigenfunctions, which means that we ignore the contributions coming from the mixing with the others representations. We then obtain the quasi-eigenvalue function $\lambda(\gamma, n,\ep)$ of the bare kernel relative to this representation:
\be
\label{kernbn}
\hat K \cdot (\vec \k^2)^{\gamma -1- \frac{n}{2}}(\vec b . \vec \k)^n= \lambda (\gamma,n, \ep) \; (\vec \k^2 )^{\gamma -1 - \frac{n}{2} + \ep} (\vec b . \vec \k)^n
\ee
with
\be
\label{lambdagamn}
\lambda(\gamma, n,\ep) = 
 \frac{1}{(4 \pi)^{\ep}} \left [ b(\gamma,n, \ep) - \half b(0,0, \ep) \right ] 
\ee 
where
\bea
\label{bgamman} 
 b(\gamma,n, \ep) = \frac{1}{\Gamma (\ep) } 
B(\ep, 1 + \frac{n}{2} -\gamma -\ep) \ B(\ep, \frac{n}{2} + \gamma + \ep) \;.
\eea 
As for the $n=0$ case we define 
\bea
\label{lambda1n}
\lambda_{1} (\gamma,n, \ep) &=& \left(\gamma+\frac{n}{2}\right) \lambda (\gamma-\ep,n, \ep) 
\\
&=&  \frac{1}{(4 \pi)^{\ep}} \Gamma (1+ \ep)  \left[ 
\frac {\Gamma (1+\frac{n}{2} - \gamma )}{\Gamma (1+\frac{n}{2}-\gamma + \ep)} 
\frac {\Gamma (1+\frac{n}{2}+ \gamma )}{\Gamma (1+\frac{n}{2}+ \gamma +  \ep)} \right. \nonumber
\\
&+& \left. \frac{\gamma+\frac{n}{2}}{\ep} \left ( \frac {\Gamma (1+\frac{n}{2} - \gamma )}{\Gamma(1+\frac{n}{2}-\gamma + \ep)} 
\frac {\Gamma (1+\frac{n}{2}+ \gamma )}{\Gamma (1+\frac{n}{2}+ \gamma +  \ep)} -
\frac {\Gamma (1  + \ep) \Gamma (1  - \ep)}{\Gamma (1 + 2\ep)} \right ) \right] \nonumber \;.
\eea
We expand in $\ep $ eq.(\ref{lambda1n}) up to the first power and write the result in the $\overline {MS}$ renormalization scheme
\be
\label{lambda1nexp}
\lambda_{1} (\gamma,n, \ep)= \lambda_{1} (\gamma,n, 0) + \ep  \; \lambda_{1}^{(\ep)} (\gamma,n) + {\cal O} (\ep^2)\;.
\ee
Where $\lambda_{1} (\gamma,n, 0)$ coincides with  the known one-loop BFKL eigenvalue function $\chi_n$ for the corresponding $\{\gamma,n\}$ representation,
\bea
\label{lambda10nchi}
 \chi_n (\gamma) &=& \frac{1}{\gamma +\frac{n}{2}} \ \lambda_{1} (\gamma,n, 0) \\
&=& 2 \psi (1) - \psi(1+\frac{n}{2}-\gamma) - \psi (\frac{n}{2}+\gamma) \nonumber
\eea
and  the contribution proportional to the firstpower of $\ep$ is given  by
\be
\label{lambda1neps}
\lambda_{1}^{(\ep)} (\gamma,n) = \frac{\gamma+\frac{n}{2}}{2} \left (2 \psi\p(1)  - \psi\p(1+\frac{n}{2}-\gamma) - \psi\p (\frac{n}{2}+\gamma)  +  \chi_n^2 (\gamma)\right) \;.
\ee
Like in the $n=0$ case, we also write  the expansion in powers of $\gamma$ of these  functions:
\be
\label{lambda1n0}
 \lambda_{1} (\gamma, n,0) = 1 + \left[\psi (1)- \psi (1+n)\right] \left(\gamma + \frac{n}{2}\right)
$$ $$
+ \sum_{k=1}^{\infty} \left[ \zeta (2 k, 1+n) - \zeta (2 k)\right] \left(\gamma + \frac{n}{2}\right)^{2 k}
$$ $$
+ \sum_{k=1}^{\infty} \left[ \zeta (2 k+1, 1+n) + \zeta (2 k+1)\right] \left(\gamma + \frac{n}{2}\right)^{2 k+1}
\ee
and 
\be
\label{lambda1nepsexp}
\lambda_{1}^{(\ep)} (\gamma,n) =\psi (1)-  \psi (1+n)
$$ $$
+\left[\frac{1}{2} \left(\psi^2(1+n)-\psi^{2}(1)+ \psi\p(1+n)-\frac{\pi
   ^2}{6} \right)-\psi (1)  \left(\psi(1+n)-\psi (1) \right)\right] \left(\gamma + \frac{n}{2}\right)
$$ $$
+\left[2 \zeta(3)+\left(\psi(1+n)-\psi (1)\right ) \left(\frac{\pi
   ^2}{6}-\psi\p(1+n)\right)     \right] \left(\gamma + \frac{n}{2}\right)^2
$$ $$
 +\sum _{p=1}^{\infty}  \left[ ( \psi (1)-\psi(1+n))( \zeta (2 p+1) +\zeta (2 p+1,1+n)) -(2 p+1) \zeta (2 p+2) + \left(p-\frac{1}{2}\right) \right.
$$ $$
\left. \times H_n^{(2 p+ 2)} + 2 \sum _{k=1}^{p-1}  \zeta (2 k+1 ) \zeta (2 (p-k)+1,1+n ) + \frac{1}{2} \sum _{k=1}^{2 p-1} H_n^{(k+1)} H_n^{(2 p-k+1)} \right] \left(\gamma + \frac{n}{2}\right)^{2 p+1}
$$ $$
+\sum _{p=2}^{\infty} \left[2 \zeta (2 p+1)+ (\psi (1+n) -\psi (1)) H_n^{(2 p)} + (p-1) H_n^{(2 p+1)}  \right.
$$ $$
 \left.-  \sum _{k=1}^{p-1} \left[\zeta (2 k+1) + \zeta (2 k+1,1+n) \right] H_n^{(2 ( p-k))} \right] \left(\gamma + \frac{n}{2}\right)^{2 p}
\ee
where we have used the  Hurwitz Zeta function (given here for a positive real number $a$, since $n$ is always positive)
\be
\zeta (s,a )= \sum_{k=1}^{\infty}\frac{1}{(k+a-1)^s}
\ee
and the  harmonic numbers of order $r$ given by
\be
H_n^{(r)}= \sum_{k=1}^{n}\frac{1}{k^r} \; .
\ee
Note that  
 we merely recover the  expressions (\ref{lambda10exp}) from (\ref{lambda1n0}) and (\ref{lambda1epsexp}) from (\ref{lambda1nepsexp}) in the $n=0$ case, since $\zeta (k,1 )=\zeta (k )$.

\section{The BFKL anomalous dimension $\gamma_{\o}(n,\ep)$}
\setcounter{equation}{0}

\subsection{Case $n=0$}
We now look at the solution $\hat \gamma_{\o} (\ep)$  of the operatorial equation
\be
\label{gammahat}
\hat \gamma_{\o} (\ep)= \lambda_1 (\hat \gamma_{\o}(\ep), \ep) \hat \gamma_{\o}^{(0)} \;.
\ee
Doing  the expansion of $ \lambda_1$ in power of $\gamma$
\be
\label{lamda1hat}
 \lambda_1 (\gamma, \ep) = \sum_{N=0}^{\infty} a_N(\ep) \gamma^N \; ,
\ee
 the coefficients $a_N (\ep) $ are known  from (\ref{lambda1ep}), (\ref{lambda10exp}) and (\ref{lambda1epsexp}), in particular $a_0 (\ep)= 1$. Therefore we can get $\hat\gamma_{\o}$ as a power series in the operator $\hat \gamma_{\o}^{(0)} $
\be
\label{gammasol}
\hat \gamma_{\o} (\hat \gamma_{\o}^{(0)},\ep) = \sum_{N=0}^{\infty} b_N(\ep) 
(\hat \gamma_{\o}^{(0)})^{N+1} 
\ee
and  we know already that $b_0 (\ep)= a_0 (\ep) $. We find the iterative relation
\be
\label{b}
 b_{N}(\ep) = \sum_{k=1}^N   a_{k}(\ep) c_{N-k}^{(k)}(\ep) 
\ee
where
\be
c_{0}^{(n)}(\ep) = (b_0(\ep))^n
\ee
and 
\be
\label{ccoeff}
c_{m}^{(n)} (\ep)= \frac{1}{m b_0(\ep)} \sum_{k=1}^m \left( k (1+n)-m \right)  b_k(\ep) c_{m-k}^{(n)}(\ep)
\ee
for $m \geq 1$.
Finally we shall need this solution up to first order in 
$\ep$ and with the shift operator removed, 
$\hat \gamma_{\o}^{(0)} \rightarrow \gamma_{\o}^{(0)}= \frac{\bar\alpha_S }{\o } $.
We then can write 
\be
\label{gammaexp}
\gamma_{\o} (\gamma_{\o}^{(0)},\ep) =
\gamma_{\o}(\gamma_{\o}^{(0)}) + \ep \; \gamma_{\o}^{ (\ep)} (\gamma_{\o}^{(0)}) + 
  {\cal O} (\ep^2)
\ee
where these quantities can be expanded  in power series 
of $\gamma_{\o}^{(0)}$ as
\be
\gamma_{\o} (\gamma_{\o}^{(0)})= \sum_0^{\infty} b_N(0) ( \gamma_{\o}^{(0)})^{N +1}
\ee
and
\be
\gamma_{\o}^{ (\ep)}(\gamma_{\o}^{(0)}) =\sum_0^{\infty} b^{ (\ep)}_N(0) ( \gamma_{\o}^{(0)})^{N +1} 
\ee
with the following coefficients 
\be
b_N(\ep) =
 b_N(0) + \ep \;  b_N^{ (\ep)}(0)   
+  {\cal O} (\ep^2)
\ee
obtained through the previous relation (\ref{b}).
We  give  the first terms of these series:
\be
\label{BFKLanomdim}
 \gamma_{\o} (\gamma_{\o}^{(0)})= \gamma_{\o}^{(0)} + 2 \zeta(3) \left( \gamma_{\o}^{(0)} \right)^4 + 2 \zeta(5) \left( \gamma_{\o}^{(0)} \right)^6 + 12 \zeta^2(3) \left( \gamma_{\o}^{(0)} \right)^7 + {\cal O} \left( (\gamma_{\o}^{(0)})^8 \right)
\ee
which corresponds to  the well-known BFKL anomalous dimension in the $n=0$ 
gluon channel and to its $\ep$-corrections: 
\bea
\label{BFKLanomdimeps}
 \gamma_{\o}^{ (\ep)} (\gamma_{\o}^{(0)})&=&  2 \zeta(3) \left( \gamma_{\o}^{(0)} \right)^3 -\; 3 \zeta(4) \left( \gamma_{\o}^{(0)} \right)^4 +\; 2 \zeta(5) \left( \gamma_{\o}^{(0)} \right)^5 \nonumber
 \\
 &+& \left[22 \zeta^2(3)- 5 \zeta(6) \right] \left( \gamma_{\o}^{(0)} \right)^6 + {\cal O} \left( (\gamma_{\o}^{(0)})^7 \right) \, \,  \, \, \, \,. 
\eea
Note that $\gamma_{\o}$, can directly be obtained from the  order $\ep^0$ of eq.(\ref{gammahat}) (with shift operator replaced by unity) as the solution of 
the usual implicit equation 
\be
 1 = \gamma_{\o}^{(0)} \chi(\gamma_{\o})
\ee
 and the $\ep$-correction $\gamma_{\o}^{ (\ep)}$  is  given from the  order $\ep$ terms  by 
\be
\label{eqgammaeps}
 \gamma_{\o}^{ (\ep)} = \lambda_1^{(\ep)}\gamma_{\o}^{(0)} + \gamma_{\o}^{ (\ep)} \left[1 + \gamma_{\o}^{(0)} \gamma_{\o} 
\chi\p (\gamma_{\o})\right]
\ee
where we have used the relation
\be
\label{derlambda1}
 \gamma_{\o}^{(0)} \lambda_1\p (\gamma_{\o}, 0) =
 1 + \gamma_{\o}^{(0)} \gamma_{\o} 
\chi\p (\gamma_{\o}),
\ee
which  leads to the solution
\be
\label{gammaeps}
 \gamma_{\o}^{ (\ep)} = \frac{\lambda_1^{(\ep)} (\gamma_{\o})}{-\gamma_{\o} \chi\p(\gamma_{\o})} \; 
\ee
with $\lambda_1^{(\ep)}$ defined in (\ref{lambda1eps}).\\
We now consider the correcting terms  to the previous eq.(\ref{eqgammaeps}) coming from the running of the  dimensionless strong coupling as $\bar\alpha_S^{   r  } =\bar \alpha_S \;(\frac{\mu_R^2}{\mu^2})^{\ep}$: it leads to write 
\bea
 \gamma_{\o}^{ (\ep)}-\ln(\frac{\mu_R^2}{\mu^2}) \frac{\chi(\gamma_{\o})}{\chi\p(\gamma_{\o})} &=& \lambda_1^{(\ep)}\gamma_{\o}^{(0)} + \gamma_{\o}^{ (\ep)} \left[1 + \gamma_{\o}^{(0)} \gamma_{\o} 
\chi\p (\gamma_{\o})\right] +\ln(\frac{\mu_R^2}{\mu^2}) \left[-\frac{\chi(\gamma_{\o})}{\chi\p(\gamma_{\o})}-\gamma_{\o} \nonumber \right.  \\
&& \left. + \lambda_1 ( \gamma_{\o}, 0)  \gamma_{\o}^{(0)}\right] \; ,
\eea
where we have used the relation
\be
 {{\partial   \gamma_{\o} } \over { \partial \ln \as}}
= - { {\chi (\gamma_{\o}) } \over {   \chi^{\prime} (\gamma_{\o}) }} \;.
\ee
These terms explicitly cancel, thus restoring the original equation (\ref{eqgammaeps}), and leading to the same expression of $\gamma_{\o}^{ (\ep)}$. Therefore, the renormalization scale dependence of the dimensionless strong coupling does not modify the solutions $(\gamma_{\o},\gamma_{\o}^{ (\ep)})$ of the  equation (\ref{gammahat}), which legitimates to neglect it during the calculation in our approach. Note that the same cancellation occurs in the Catani-Hautmann calculation if such running effects are considered, then eq.(B.15) (from which the $R_N$ factor is obtained) in \cite{CH} will be unchanged.

\subsection{Case $n> 0$}

We now look at the solution of the operatorial equation in the 
general conformal spin $n$ case
\be
\label{gammahatn}
 \hat \gamma_{\o}(n,\ep) = \lambda_1 \left(\hat \gamma_{\o}(n,\ep)-\frac{n}{2}, \ep \right) \hat \gamma_{\o}^{(0)}
\ee
whose   expansion in power of $\hat\gamma_{\o}^{(0)}$ can easily be obtained from eqs.(\ref{lambda1nexp}), (\ref{lambda1n0}) and  (\ref{lambda1nepsexp}) following the same procedure as previously for $n=0$, see eqs.(\ref{lamda1hat}) to (\ref{ccoeff}). After replacement of the shift operator by unity, we expand the solution  in $\ep$ as 
\be
\gamma_{\o} (\gamma_{\o}^{(0)},n,\ep) =
\gamma_{\o}(\gamma_{\o}^{(0)},n) + \ep \; \gamma_{\o}^{(\ep)} (\gamma_{\o}^{(0)},n) + 
  {\cal O} (\ep^2) \;.
\ee
 The equation (\ref{gammahatn}) writes now  at ${\cal O} (\ep^0)$   accuracy
\be
\gamma_{\o}(n)= \lambda_1 \left(\gamma_{\o} - \frac{n}{2} ,n, 0 \right) \gamma_{\o}^{(0)}
\;,
\ee
which leads with the use of the relation (\ref{lambda10nchi}), to the 
implicit equation defining  the BFKL anomalous dimension $\gamma_{\o}(n)$  in the general conformal spin $n$ gluon channel
\be
\label{BFKLanomdimn}
 1 = \gamma_{\o}^{(0)} \chi_n \left(\gamma_{\o}(n) - \frac{n}{2} \right) \;.
\ee
The result of the expansion of this quantity in power series of $\gamma_{\o}^{(0)}$ reads now for the firsts coefficients
\bea
\label{BFKLanomdimnexp}
 \gamma_{\o} (\gamma_{\o}^{(0)},n) &=&  \gamma_{\o}^{(0)}  + \;  \left[ \psi (1) - \psi(1+n) \right] \left( \gamma_{\o}^{(0)} \right)^2  \nonumber
\\
&&  \hspace{-2cm} +  \; \left[  \left(\psi (1) - \psi(1+n) \right)^2 +  \zeta(2,1+n)-  \zeta(2)  \right] \left( \gamma_{\o}^{(0)} \right)^3  + {\cal O} \left( (\gamma_{\o}^{(0)})^4 \right) \;.
 \eea
The ${\cal O} (\ep)$ terms of the equation (\ref{gammahatn}) lead to
the $\ep$-correction $\gamma_{\o}^{(\ep)}(n)$ of the  BFKL anomalous dimension  with conformal spin $n$
\be
\label{gammaneps}
 \gamma_{\o}^{(\ep)}(n)= \frac{\lambda_1^{(\ep)} (\gamma_{\o}(n) - \frac{n}{2},n)}{-\gamma_{\o}(n) \; \chi_n\p(\gamma_{\o}(n) - \frac{n}{2})} \; 
\ee
with $\lambda_1^{(\ep)}(\gamma,n)$ defined in the  eq.(\ref{lambda1neps}). Its perturbative expansion reads 
\bea
\label{BFKLanomdimepsnexp}
 \gamma_{\o}^{ (\ep)} (\gamma_{\o}^{(0)},n) =    \left[\psi (1) - \psi(1+n) \right]  \gamma_{\o}^{(0)}  +  \; \left[  \frac{3}{2} \left(\psi (1) - \psi(1+n) \right)^2 \right. \nonumber
\\
\left. +  \frac{1}{2}  \left(\zeta(2,1+n)-  \zeta(2) \right)   \right] \left( \gamma_{\o}^{(0)} \right)^2  + {\cal O} \left( (\gamma_{\o}^{(0)})^3 \right) \;.
\eea
Whereas the  Riemann Zeta function $\zeta(k)$ is evaluated at natural numbers $k$ (related to the order of the perturbation) in the coefficients of the series expansion  (\ref{BFKLanomdim}) and  (\ref{BFKLanomdimeps}) for the $n=0$ case, it also appears as Hurwitz Zeta function   $\zeta(k,1+n)$ in  the $n\neq0$ case.  We also notice that the NLO and NNLO coefficients of the BFKL anomalous dimension are no more vanishing for the general $n$ case.

\section{The treatment of shift operator}
\setcounter{equation}{0}

For $\lambda_1 = 1$ (\ref{F<}) simplifies to
\be
\label{F0<}
F_{0 <} (\o, \k) = \bar \alpha_S (\frac{\mu_F^2}{\mu^2})^{\ep}
\int d\gamma (\frac{\k}{\mu_F^2})^{\gamma}
\frac{1}{ \o - \bar \alpha_S  e^{-\ep \dd_{\gamma} } 
\frac{1}{\gamma + \ep } }
\frac{1}{\ep- \gamma}  
\ee
The pole in the last factor lies to the right of the contour.\\

In the first way calculation we decompose the factor
involving the shift operator in powers of $\bar \alpha_S $
before doing the $\gamma $ integral. The the Nth term
in the integrand
is
\be \frac{1}{\o} 
\left (\frac{1}{\gamma} \hat \gamma_{\o}^{(0)} \right )^N
\frac{1}{\ep-\gamma} \;.
\ee
We should move $\hat \gamma_{\o}^{(0)}$ to one side.
Here it is not convenient moving to the right
because the pole of the right-most factor should be kept
to the right of the contour. We conjugate the operator
$(\frac{1}{\gamma} \hat \gamma_{\o}^{(0)})^N$ so that now 
the shift operators are acting to the left and then we 
move  the shift operators to the left.
\be \frac{1}{\o}
\left ( \hat \gamma_{\o}^{(0)\dagger } \right )^N
\frac{1}{\gamma + (N-1)\ep} \ \frac{1}{\gamma + (N-2)\ep}
...\frac{1}{\gamma + \ep} \ \frac{1}{\gamma }
\ \ \frac{1}{\ep-\gamma} \;.
\ee
We calculate the residues in the poles to the left of the 
contour and replace the shift operators by unity 
\be \frac{1}{\o}
\left (  \gamma_{\o}^{(0)} \right )^N
\ep^{-N} \sum_0^{N-1} (-1)^k \frac{1}{(k+1)!} 
\frac{1}{(N-k-1)!} \ee
The sum can be written as
\be \frac{1}{(N-1)!} \int_0^1 dx (1-x)^{N-1} = \frac{1}{N!} \;. \ee
Notice that the contribution of $N=0$ vanishes, therefore
the result is 
\be \frac{1}{\o} \left 
( \exp (\frac{1}{\ep} \gamma_{\o}^{(0)} ) -1 \right ).
\ee
As the alternative way of calculation we would like to
do the $\gamma$ integral first. 
Write the integral in (\ref{F0<}) as
\be
\int d\gamma \frac{1}{\gamma - \hat \gamma_{\o}^{(0)} }
\ \frac{\gamma}{\o} \ \frac{1}{\ep- \gamma} \;.
\ee
In order to release the integration variable from its
role of the operator canonically conjugated to 
$-i \dd_{\gamma}$ we substitute $\gamma $ by 
$\gamma + \gamma\p$ in the factors besides 
of the pole factor and also $-i \dd_{\gamma}$ 
by $-i \dd_{\gamma}\p$. We should set $\gamma\p = 0$
after all differentiations have been done.
The residue is
\be \left. \frac{\hat \gamma_{\o}^{(0)} + \gamma\p}{\o} \ 
\frac{1}{\ep- \hat \gamma_{\o}^{(0)} + \gamma\p} 
\right |_{\gamma\p = 0} \;.
\ee
We write the second factor as
\be \frac{1}{1- \frac{1}{-\gamma\p +\ep} 
\hat \gamma_{\o}^{(0)} } \  \frac{1}{-\gamma\p +\ep}
\ee 
and expand in powers of $\bar \alpha_S $. The Nth term
reads 
\be \frac{\hat \gamma_{\o}^{(0)} + \gamma\p}{\o}
\left (\frac{1}{-\gamma\p +\ep} \hat \gamma_{\o}^{(0)} 
\right )^N \frac{1}{-\gamma\p +\ep} \;.
\ee
Now we move the operators acting on $\gamma\p$ to the right.
After this the shift operators can be replaced by unity and we get
\be \frac{1}{\o}\ 
\frac{1}{-\gamma\p +2\ep} \ \frac{1}{-\gamma\p +3\ep} ...
   \frac{1}{-\gamma\p +N \ep} \ 
\frac{1}{-\gamma\p + (N+1)\ep}
 \ (\gamma_{\o}^{(0) })^{N +1} + \frac{\gamma\p}{\o} ...  
\gamma_{\o}^{(0) N } \;.
\ee
We are advised to set $\gamma\p = 0$, 
we  obtain 
\be
\frac{1}{\o} (\ep)^{-N-1} \frac{1}{(N+1)!} 
(\gamma_{\o}^{(0)})^{ N +1} 
\ee
and the sum over $N$ yields the same result as
in the first approach.\\

Preparing for the discrete sum case 
the treatment of a non-trivial
$\lambda_1(\gamma, \ep)$  we
outline the alternative way
of treating  (\ref{CHeps}).\\
The condition $\gamma = 0$ is implemented by contour
integral and  then the contour is deformed 
to enclose the anomalous dimension instead of
the origin in $\gamma$ plane.
\bea
 \left. \frac{1}{1- \hat \gamma_{\omega}^{(0)} 
\frac{1}{\gamma} } \right|_{\gamma = 0} &=&
\frac{1}{2\pi i} \oint_{C_0} \frac{d\gamma}{\gamma} 
\frac{1}{1- \hat \gamma_{\omega}^{(0)} 
\frac{1}{\gamma} } = 
\left.
\frac{1}{2\pi i} \oint_{C_0} \frac{d\gamma}{\gamma} 
\frac{1}{1- \hat \gamma_{\omega}^{(0) \prime} 
\frac{1}{\gamma + \gamma\p} } 
\right|_{\gamma\p = 0}  \nonumber \\
&=& \left.
1 + \frac{1}{2\pi i} \oint_{C_0} \frac{d\gamma}{\gamma}
\hat \gamma_{\omega}^{(0) \prime} 
\frac{1}{\gamma + \gamma\p- 
\hat \gamma_{\omega}^{(0) \prime}  } 
\right|_{\gamma\p = 0} \;.
\eea
The shift operator is now acting on $\gamma\p$ and
is disentangled from the integration variable.
The subtraction removes the singularity at 
$\gamma = \infty$ in the integrand. After this the
contour $C_0$ can be deformed to $\hat C$ encircling the pole of 
the second factor (with opposite orientation).
It is convenient to change the integration 
variable to 
$\gamma^{\prime \prime} = \gamma + \gamma\p$.
This is accompanied by a transposition of the operators
acting now to the left.
We evaluate the integral taking residue
 \bea
 \left. \frac{1}{1- \hat \gamma_{\omega}^{(0)} 
\frac{1}{\gamma} } \right|_{\gamma = 0} &=& \left. 1 
-\frac{1}{2\pi i} \oint_{\hat C }  
\frac{d\gamma^{\prime \prime}}
{\gamma^{\prime \prime} - \gamma\p} 
\frac{1}{\gamma^{\prime \prime}- 
\hat \gamma_{\omega}^{(0) \prime T}  }
\hat \gamma_{\omega}^{(0) \prime T} 
\right|_{\gamma\p = 0} \nonumber \\
&=& 1+ \frac{1}{\gamma^{\prime }- 
\hat \gamma_{\omega}^{(0) \prime T}  } \left.
\hat \gamma_{\omega}^{(0) \prime T} \right|_{\gamma\p = 0}
= 
1+ \sum_{N=0}^{\infty} \left.
\frac{1}{\gamma\p} \hat \gamma_{\omega}^{(0) \prime T} 
\frac{1}{\gamma\p} \hat \gamma_{\omega}^{(0) \prime T} 
...
\frac{1}{\gamma\p} \hat \gamma_{\omega}^{(0) \prime T} 
\right|_{\gamma\p = 0} \nonumber \\
&=& 1+ \sum_{N=0}^{\infty} 
\gamma_{\omega}^{(0)\ N+1} 
\frac{1}{ (N+1)! \ep^{N+1}}
= e^{\frac{\gamma_{\omega}^{(0)}}{\ep} } \;.
\eea

\section{The asymptotics $\ep \to 0$}
\setcounter{equation}{0}

Let us  write the integral  appearing in the 
results (\ref{Fresult1}) and (\ref{Fresult2}) as
\be
 I(\ep) = \frac{1}{\ep}\int_0^1  \frac{d\beta}{\beta}
\gamma(\beta) \ A(\beta) \exp(\frac{1}{\ep} \int_{\beta}^1  
\frac{d\beta_1}{\beta_1} \gamma(\beta) )  \;.
\ee
We change the integration variable to 
$y(\beta) = \int_{\beta}^1 \frac{d\beta_1}{\beta_1} 
\gamma(\beta_1) $ and denote the resulting function 
from this substitution
by $\tilde A(y) = A(\beta (y) ) $ and also $y_0 = y(0)$.
Notice that the value $y=0$ corresponds to $\beta = 1$. 
\bea
 I(\ep) &=& \frac{1}{\ep} \int_0^{y_0} dy \tilde A(y) 
\exp (\frac{1}{\ep} y)
= \frac{y_0}{\ep} \int_0^1 dz \tilde A(y_0 z) \exp (\frac{y_0}{\ep}  z) \nonumber \\
&=& y_0 \int_0^{\ep^{-1}} dz_1 \tilde A(\ep y_0 z_1) \exp (y_0 z_1) \;.
\eea
We apply the mean value theorem and obtain
\bea 
I(\ep)&=&  y_0 \tilde A(\ep y_0 z_0) \int_0^{\ep^{-1}} dz_1  \exp (y_0 z_1) \nonumber \\
&=&  \tilde A(\ep y_0 z_0) ( \exp (\frac{1}{\ep} y_0 ) -1 )
\eea
for some $z_0$ in the integration range. The exponential function in the
integrand ensures that $y_0 z_0 \sim 1 $ is independent of $\ep$. 
We have noticed above that at the argument $y=0$ of the function 
$\tilde A(y)$ is equal to the function $A(\beta)$ at $\beta=1$
Therefore,
the asymptotics  in $\ep$ is
\be I(\ep) = 
A(1) \left(\exp \left(\frac{1}{\ep} \int_0^1  \frac{d\beta}{\beta} \gamma(\beta) \right) -1\right) 
(1+{\cal O}(\ep) )\;.
\ee
\\

\end{document}